\documentclass[11pt,a4paper]{article}

\usepackage{url} 
\usepackage{soul}
\usepackage[T1]{fontenc}
\usepackage{graphicx}
\usepackage{amsmath,amssymb,amsfonts}
\usepackage{bm}
\usepackage{xcolor}
\usepackage{natbib}
\usepackage[a4paper,margin=2.5cm]{geometry}
\usepackage[colorlinks=true,linkcolor=blue,citecolor=blue,urlcolor=blue,breaklinks]{hyperref}

\newcommand{\bs}{\bm}
\newcommand{\mr}{\mathrm}

\title{A discrete adjoint method for deterministic and probabilistic eikonal-equation-based inversion of traveltime for velocity and source location}

\date{}
\author{Andrea Zunino$^{1}$\thanks{corresponding author: andrea.zunino@eaps.ethz.ch},
  Scott Keating$^{1}$, Andreas Fichtner$^{1}$ \\
 $^1$ Department of Earth and Planetary Sciences, ETH Zurich, Switzerland }

\graphicspath{{./figs/}}

\begin{document}

\maketitle

\begin{abstract}
  Seismic traveltime tomography represents a popular and useful tool for unravelling the structure of the subsurface across the scales. In this work we address the case where the forward model is represented by the eikonal equation and derive a formalism to solve the inverse problem where gradients are calculated efficiently using the discrete adjoint state method.
Our approach provides gradients with respect to both velocity structure and source locations, allowing us to perform a consistent joint inversion. The forward problem is solved using a second-order fast-marching method, which provides a strategy to efficiently solve the adjoint problem. Our approach allows for arbitrary positions of both sources and receivers and for a refined grid around the source region to reduce errors in computed traveltimes. We show how gradients computed using the discrete adjoint method can be employed to perform either deterministic inversion, i.e., solving an optimization problem, or for a probabilistic (Bayesian) approach, i.e., obtaining a posterior probability density function. We show applications of our methodology on a set of synthetic examples both in 2D and 3D using the L-BFGS algorithm for the deterministic case and the Hamiltonian Monte Carlo algorithm for the probabilistic case.
\end{abstract}


\section{Introduction}
\label{sec:intro}

Seismic traveltime tomography has proven to be one of the most successful tools to investigate the structure of the subsurface for decades \citep[e.g.,][]{akiDeterminationThreedimensionalVelocity1976,noletBreviarySeismicTomography2008,rawlinsonSeismicTomographyWindow2010}. This approach has been and continues to be used successfully across the scales from local exploration to global settings \citep[e.g.,][]{thurberEarthquakeLocationsThreedimensional1983,husenLocalEarthquakeTomography2001,simmonsLLNLG3Dv3GlobalWave2012}, despite the rise of the full-waveform inversion (FWI) strategy in the past thirty years \citep[e.g.,][]{bunksMultiscaleSeismicWaveform1995,gerhardprattGaussNewtonFullNewton1998,chararaFullWaveformInversion2000,trompSeismicTomographyAdjoint2004,Brossier_2009a,tapeAdjointTomographySouthern2009,virieuxOverviewFullwaveformInversion2009,fichtnerFullSeismicWaveform2011,bozdagGlobalAdjointTomography2016}. The latter is capable of exploiting almost completely the physics of wave propagation, producing spectacular results; however, the computational burden associated with FWI is high and hence there are several cases where traveltime tomography still finds applications.
A widespread approach for modeling first-arrival traveltimes under the assumption of infinite frequency is to employ ray theory \citep[e.g.][]{virieuxRayTracing3D1991,cervenySeismicRayTheory2001,noletBreviarySeismicTomography2008}. The resulting forward problem is non-linear as the arrival time depends not only on the velocity of the medium but also on the path followed by the ray. The solution to the inverse problem is then typically based upon a linearization which amounts to two steps: i) tracing rays in a prior model and ii) updating the velocity model to improve the fit to observed data \citep[e.g.,][]{thurberEarthquakeLocationsThreedimensional1983,thurberHypocentervelocityStructureCoupling1992}. Unfortunately, ray tracing techniques suffer from lack of robustness (failure to converge) in the case of complex velocity models \citep{cervenySeismicRayTheory2001}. A different, ray-free strategy is that of grid-based finite-difference solution of the eikonal equation \citep[e.g.,][]{vidaleFinitedifferenceCalculationTravel1988,vidaleFinitedifferenceCalculationTraveltimes1990,podvinFiniteDifferenceComputation1991,picaFastAccurateFinite1997,qianAdaptiveFiniteDifference2002,rawlinsonWaveFrontEvolution2004,zhangEikonalSolverCelerity2005,fomelFastSweepingMethod2009,luoHigherorderSchemes3D2012a,nobleAccurate3DFinite2014,lebouteillerDiscontinuousGalerkinFastsweeping2019}. In this approach, the velocity model is discretised over a grid, and the traveltime is computed by finite differences at all grid points and finally extracted at the receiver locations. In this way, contrary to the ray-based approach, which requires the calculation of each source-receiver path separately, the traveltimes at all receivers for a given source are obtained simultaneously. The other important advantage over the ray-based approach is the possibility to avoid linearisation when solving the inverse problem, as discussed below.

The first grid-based solvers for the forward problem were based on an expanding box technique, where  updates of the traveltime from the source location outwards was performed by using a finite-difference scheme  \citep{vidaleFinitedifferenceCalculationTravel1988,vidaleFinitedifferenceCalculationTraveltimes1990,trierUpwindFinitedifferenceCalculation1991}. \citet{podvinFiniteDifferenceComputation1991} developed a related, massively parallel algorithm, also incorporating sharp contrasts in velocity.
Later, the fast marching method (FMM) was developed for the general problem of tracking the evolution of interfaces \citep{sethianFastMarchingLevel1996a,sethianAdvancesFastMarching1999,sethianFastMarchingMethods1999}. The FMM soon found application to seismic problems \citep[e.g.,][]{sethian3DTraveltimeComputation1999,rawlinsonMultipleReflectionTransmission2004,rawlinsonWaveFrontEvolution2004}. The main advantages of FMM over previously proposed methods were improved computational efficiency and the capability to handle complex velocity models \citep{sethian3DTraveltimeComputation1999,rawlinsonWaveFrontEvolution2004}.

Inaccurate traveltime values near the source position, due to the high curvature of the wavefront, are the major source of error for grid-based methods. To reduce these errors, the FMM allows for a refined grid around the source location, improving the accuracy of the traveltime computed there \citep[e.g.,][]{rawlinsonWaveFrontEvolution2004}. The traveltimes obtained in the fine grid can then be mapped onto the coarse grid, where the rest of the computations are performed. Another approach is that of employing the factored eikonal equation \citet[e.g.,][]{fomelFastSweepingMethod2009,treisterFastMarchingAlgorithm2016b}, however, the implementation becomes practically more complex.

When these approaches were first adopted for tomography, they were still using rays to solve the inverse problem \citep[e.g.,][]{zeltThreedimensionalSeismicRefraction1998,rawlinsonSeismicTraveltimeTomography2003,rawlinsonWaveFrontEvolution2004}. Specifically, the traveltimes computed by finite differences on the entire grid were used to back-trace rays from receivers to the source following the gradient of the traveltime field and, finally, these calculated rays were used to construct the sensitivity matrix. The advantage of the grid-based solvers compared to classic ray tracing lies in the fact that the cost of back-tracing rays is almost independent of the number of receivers. In other words, once the traveltime has been calculated everywhere on the grid, i.e., the most expensive task, then the rays can be cheaply back-traced from receivers to source(s) using the negative gradient of the traveltime field without any additional traveltime calculation. However, the accuracy of back-traced rays relies on the accuracy of the gradient of the traveltime and on the scheme adopted to construct the actual ray path. Few exceptions to this approach were proposed, e.g., by  \citet{lelievreInversionFirstarrivalSeismic2011}, where derivatives were obtained by explicit symbolic differentiation.

A more accurate and efficient strategy is to explicitly compute the gradient of the objective function with respect to velocity using the adjoint method as proposed by \citet[e.g.,][]{seiGradientCalculationTraveltime1994}, though this approach has been widely adopted only more recently  \citep[e.g.,][]{leungAdjointStateMethod2006a,taillandierFirstarrivalTraveltimeTomography2009,huangJointTransmissionReflection2012,liFirstbreakTraveltimeTomography2013a,liLevelsetAdjointstateMethod2014,waheedFirstarrivalTraveltimeTomography2016,tavakolib.SlopeTomographyBased2017,zuninoIntegratingGradientInformation2018,tongAdjointStateTraveltimeTomography2021,zuninoHMCLabFrameworkSolving2023a}. The adjoint state method \citep[e.g.,][]{lionsOptimalControlSystems1971,plessixReviewAdjointstateMethod2006b,hinzeOptimizationPDEConstraints2008} allows the for efficient computation of gradients, typically with a computational cost equal to that of approximately two forward calculations. The cost of computing such a gradient is practically independent of the number of receivers, making it particularly attractive for large data sets. In addition, the gradient obtained is typically accurate to machine precision.

The eikonal equation is a non-linear partial differential equation for which the adjoint equation is different from the forward model (i.e., the forward operator is not self-adjoint). Therefore, to compute gradients, an adjoint solver needs to be implemented.
The majority of works on adjoint eikonal tomography  \citep[e.g.,][]{leungAdjointStateMethod2006a,taillandierFirstarrivalTraveltimeTomography2009,huangJointTransmissionReflection2012,tongAdjointStateTraveltimeTomography2021} utilise the \emph{continuous} adjoint method: the adjoint system is derived analytically and the discretisation occurs only at the last stage of the process, i.e., the continuous adjoint equation is discretised for implementation on a computer only after being analytically determined. Thus far, relatively few works \citep[e.g.,][]{liFirstbreakTraveltimeTomography2013a,treisterFastMarchingAlgorithm2016b} employ the \emph{discrete} adjoint method, where first the forward problem is discretised and then a discrete adjoint system is derived. Such an approach presents some advantages, such as making sure that the adjoint, and hence the gradient, are exactly the couterparts of the forward model and do not suffer from potential discrepancies due to the separate discretisation of the continuous forward and adjoint systems. Another advantage is that the boundary conditions are easy to specify, since they are already defined in the discrete system.

 Several works \citep[e.g.,][]{leungAdjointStateMethod2006a,taillandierFirstarrivalTraveltimeTomography2009,huangJointTransmissionReflection2012} obtain an homogeneous, continuous adjoint equation  with a boundary condition specifying receivers located on the boundary of the domain (i.e. receivers inside the domain are prohibited). Other authors  \citep[e.g.][]{benaichoucheFirstArrivalTraveltime2015,tongAdjointStateTraveltimeTomography2021} have derived an inhomogeneous continuous adjoint equation where receivers can be located inside the domain, broadening the range of application to generic source-receiver geometries.

In addition to the problem of inversion for the structure of the subsurface, there is also that of inversion for source location. This is usually treated separately from the inversion for structure, for instance by first inverting for source location and subsequently inverting for the velocity model with sources fixed \citep[e.g.,][]{kisslingInitialReferenceModels1994}. Similarly, in certain approaches, the inversion for source location and structure are performed in an alternating fashion, by iteratively updating first one, then the other \citep[e.g.,][]{rawlinsonSeismicTomographyWindow2010}. 

Recently, \citet{tongAdjointStateTraveltimeTomography2021,tongAdjointStateTraveltimeTomography2021a,tongAdjointstateDifferentialArrival2024} have proposed a comprehensive workflow targeting most of the points discussed above. In particular, they employ the \emph{continuous} adjoint method and develop a two-step framework where first an eikonal-equation based location algorithm updates the source positions and, subsequently, the adjoint-based tomographic algorithm, based on a fast-sweeping method \citep{zhaoFastSweepingMethod2005}, updates the velocity model. 

However, a few issues remain open: i) a single, coherent, methodology that includes arbitrary position of sources and receivers, refinement of the grid around the source to improve accuracy and the use of the adjoint method to compute gradients; ii) calculation of the gradient with respect to source location using the same adjoint method as the one employed for velocity in a self-consistent manner; iii) use of the discrete adjoint method to ensure consistency between forward and adjoint solvers; iv) uncertainty quantification using a probabilistic framework for non-linear eikonal-equation-based problems.

In this paper, we develop a formalism for traveltime tomography that addresses the above-mentioned issues, providing a single coherent framework for both structure (velocity) and source location inversions. Contrary to the majority of previous works, the formulation we adopt is that of the \emph{discrete} adjoint state method, where we start by discretising both the eikonal equation and the objective function (section~\ref{sec:eiktomo}). From the discrete forward model and objective function, we derive a complete theory to compute gradients for both the structure (i.e., velocity) and source location, deriving the adjoint equation(s) directly in the discrete domain (section~\ref{sec:discradjgrad}). In this way, it is possible to perform a simultaneous inversion for both velocity and source position, making the inverse problem more self-consistent. Our forward and adjoint solvers are based on a second-order FMM method. Moreover, to ensure a high accuracy, these are derived for the cases both with and without the presence of grid refinement around the source location, as the theoretical framework is capable of accommodating such a scheme. Finally, the theoretical framework developed in this work is general with respect to the choice of objective function as well as with respect to the finite-difference-based solver used. Sources and receivers can be located anywhere in the domain, i.e., they are not restricted to coincide with grid nodes. The description of the method is developed for a Cartesian coordinate system, however, the equivalent approach in spherical coordinates on a regular grid is easy to derive by simply substituting the appropriate form of the derivatives in spherical coordinates. Our code also implements functions for spherical coordinate systems (specifically for spherical chunk models).
Moreover, we address the problem of uncertainty quantification while accounting for the full non-linearity of the problem, which, so far, has not been completely addressed in traveltime tomography, while it has been dealt with for source location. Specifically, we develop our formalism targeting both deterministic and probabilistic inversion approaches. The latter is based on the Hamiltonian Monte Carlo (HMC) method \citep[e.g.,][]{duaneHybridMonteCarlo1987,nealMCMCUsingHamiltonian2011,fichtnerHamiltonianMonteCarlo2019,zuninoHMCLabFrameworkSolving2023a}, a sampling strategy devised for non-linear inverse problems, which is capable of providing uncertainty quantification.

After developing the theoretical framework in section~\ref{sec:discradjgrad}, section~\ref{sec:numexper} illustrates a set of examples of the inverse methodology using both a deterministic approach with an L-BFGS algorithm \citep[e.g.,][]{liuLimitedMemoryBFGS1989,nocedalNumericalOptimization2006} and a probabilistic approach using the HMC algorithm \citep[e.g.,][]{nealMCMCUsingHamiltonian2011,fichtnerHamiltonianMonteCarlo2019}. The latter requires efficient computation of gradients and, contrary to the former, provides uncertainty quantification.

\section{Eikonal tomography}
\label{sec:eiktomo}

\subsection{Forward problem}
\label{sec:fwdproblem}

We are interested in the first arrival of seismic waves at the positions where receiver stations are located. The wavefront evolution, which provides the arrival time of seismic waves, is described by the eikonal equation as follows. Let $\Omega \subset \mathbb{R}^d$, ($d=2,3$) denote a domain and consider a velocity model $v = v(\bs{x})$. The eikonal equation then reads
\begin{equation}
  \label{eq:eikforwcont}
| \nabla u(\bs{x})|^2 = \sum_d \left( \dfrac{\partial u}{\partial x_d} \right)^2 = v^{-2}(\bs{x}), \quad \bs{x} \in \Omega,
\end{equation}
where $u$ is the traveltime field and $\bs{x}$ represents the position. For a source at $\bs{x}_0$ with $u(\bs{x}_0) = t_0$ and given a velocity model $v(\bs{x})$, we seek the solution $u = u(v,x)$ for the first arrival time.
There exist several algorithms to solve eq.~\eqref{eq:eikforwcont} numerically, mostly based on finite-difference approaches. Among these, two popular examples are the fast sweeping method \citep[e.g.,][]{zhaoFastSweepingMethod2005}, which solves eq.~\eqref{eq:eikforwcont} using an iterative scheme where several passes are needed for a full update of the traveltimes on the grid, and the FMM, which solves the problem in a single pass employing an upwind scheme where the wavefront is evolved in time in terms of a ``narrow band'' of grid points \citep[e.g.,][]{sethianFastMarchingLevel1996a,sethianFastMarchingMethods1999}.   

Typically, the source position is assumed to lie on a grid node. However, in this work, we allow it to be located anywhere in the domain of the model $\Omega$. Analogously, receivers can be placed anywhere in the domain. Because of this, when the source location does \emph{not} coincide with a node position in the grid, the traveltime from the source location to the enclosing grid points (4 in 2D or 8 in 3D) needs to be computed before starting the finite-difference solver. The solver uses these arrival times as the initial conditions to start the computation. The traveltime from a source not coincident with a grid point is thus computed analytically using the formula
\begin{equation}
  \label{eq:ttime_sourceregion}
  u_s = \dfrac{ \sqrt{ \sum_r (n_{sr} - s_r )^2  }}{v_s} \, ,
\end{equation}
where $u_s$ is the traveltime at a grid node surrounding the source, $ n_{sr}$ is the position of the nodes surrounding the source (e.g., in 2D, we have $s=1,..,4$, four grid points enclosing the source and $r=1,2$, two spatial coordinates), $s_{r}$ the source position and $v_s$ the velocity associated with the surrounding grid points. Other schemes based on a different kind of interpolation of the velocity are possible, however, for simplicity we will use eq.~\eqref{eq:ttime_sourceregion} hereafter.

\begin{figure}[h]
  \centering
  \includegraphics[width=0.5\textwidth]{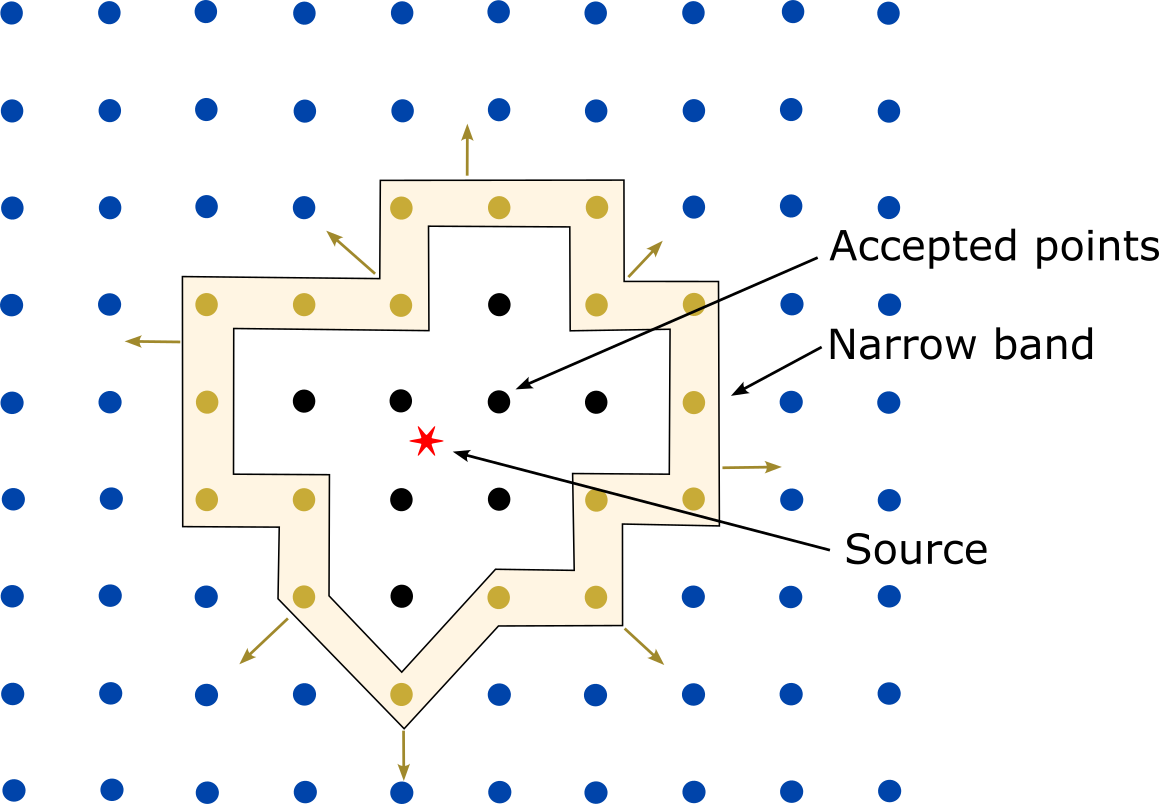}
  \caption{An example of the grid used for the FMM scheme in 2D, with arbitrary position of the source. First the traveltime at the nodes surrounding the source is computed analytically, then the fast marching algorithm is started. At each iteration, the already computed traveltimes (black grid nodes) are used to compute values at surrounding nodes. The wavefront is advanced by using traveltimes from points in the narrow band (beige grid nodes). The narrow band is then evolved until the full grid is spanned.}
  \label{fig:fmmnarrband}
\end{figure}
In this work, we employ the FMM to compute the arrival times, however, the theory illustrated in the following is general and can be applied also when using different numerical schemes to solve the forward problem. The FMM algorithm is unconditionally stable and based on a Godunov upwind discretisation \citep[e.g,][]{sethianFastMarchingLevel1996a}. Details of the implementation can be found in, e.g., \citet{sethian3DTraveltimeComputation1999} and \citet{treisterFastMarchingAlgorithm2016b}. 
The FMM is based on the fact that information propagates only from smaller to larger traveltime values (upwind scheme), hence, once the wavefront has touched a point, the value of traveltime at that point cannot be changed. The algorithm relies upon tagging the grid points with three labels, which represent points for which the traveltime is known (``known''), points for which trial traveltimes have been calculated (``front''), that constitute the ``narrow band'', and points for which the traveltime is still unknown (``unknown''). See Fig.~\ref{fig:fmmnarrband} for a visual explanation. Points belonging to the narrow band are sorted using an efficient binary heap data structure from which the point with the smaller traveltime is retrieved, tagged as ``known'' and traveltime at the neighbouring points is computed from eq.~\eqref{eq:eikgodunov} using only ``known'' points. Such points then become part of the narrow band, which approximates the shape of the wavefront. Iterating over this scheme, will eventually lead to the computation of the traveltime at all the points across the grid.

\begin{figure}[h]
    \centering
  \includegraphics[width=\textwidth]{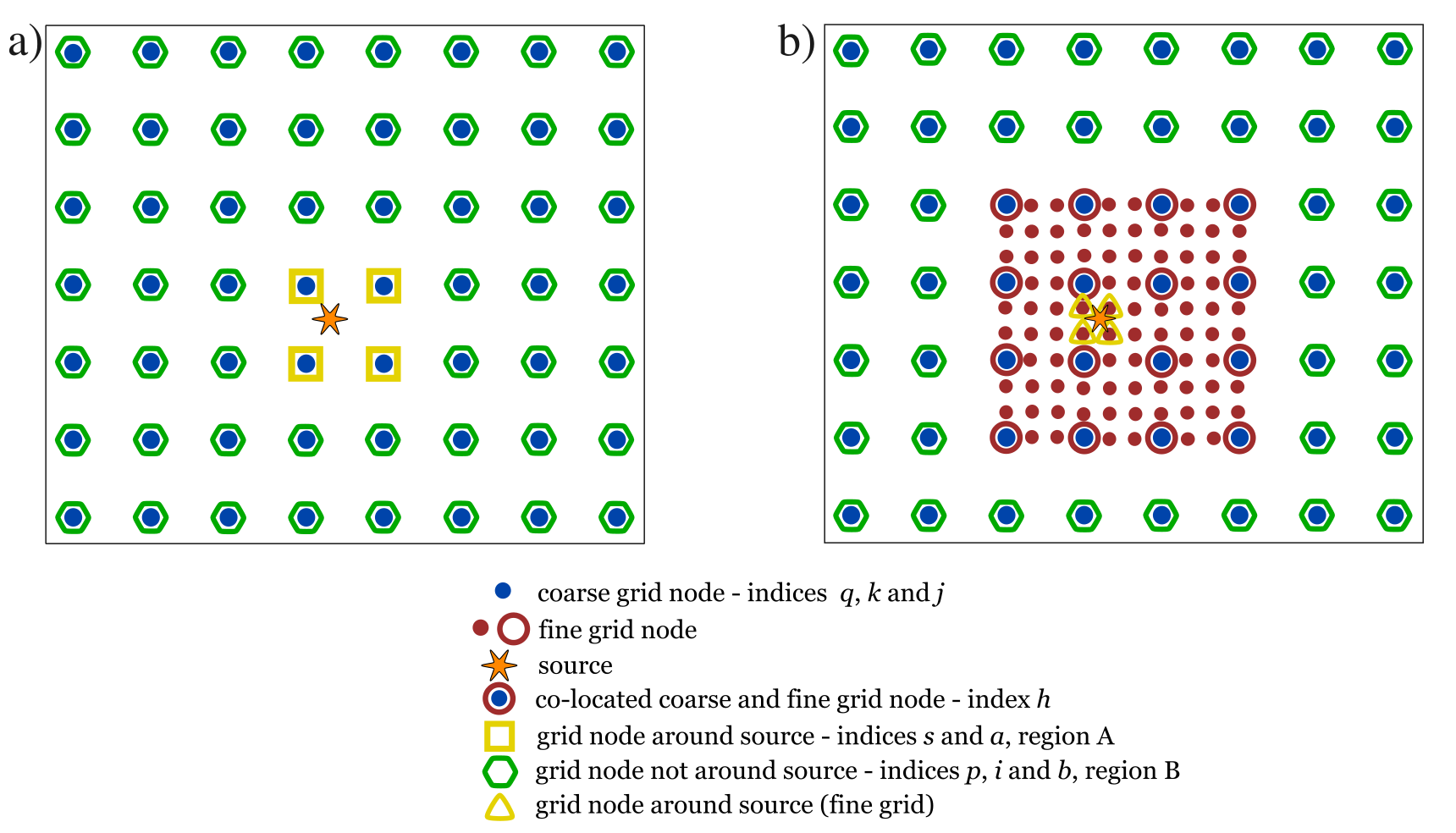}
  \caption{An example of 2D grid without (a) and with (b) grid refinement around the source. The grid spacing of the refined grid is about one third of the coarse grid. With reference to panel (a), the legend indicates also which indices belong to the various sets of points that are described in the main text. In case of grid refinement (panel (b)), visualising the sets of indices, some of which are shared between coarse and fine grids, is more difficult and therefore the reader is referred to the main text.}
  \label{fig:coarse_refined_grid}
\end{figure}

In our case, we employ a second-order FMM scheme \citep{sethianFastMarchingMethods1999,rawlinsonWaveFrontEvolution2004,treisterFastMarchingAlgorithm2016b}, where, to fulfil the upwind scheme, forward and backward formulae for derivatives are required, which read
\begin{equation}
\label{eq:fmmderiv2nd}
D^{+} t_i = \dfrac{-3 t_i + 4t_{i+1} - t_{i+2}}{2 \Delta h} \,\,  \mathrm{and} \,\,
  D^{-} t_i = \dfrac{3 t_i -4t_{i-1}+t_{i-2}}{2 \Delta h} \, ,
\end{equation}
where the superscripts $+$ and $-$ stand for forward and backward derivatives, respectively, $\Delta h$ is the grid spacing and $t$ the traveltime. In case traveltimes required for second-order stencils are not available, the algorithm resorts to first order, as given by
\begin{equation}
\label{eq:fmmderiv1st}
D^{+}  t_i = \dfrac{t_{i+1} - t_{i}}{\Delta h} \quad \mathrm{and}
\quad  D^{-} t_i = \dfrac{t_i - t_{i-1}}{\Delta h} \, .
\end{equation}
The algorithm to solve the discrete version of eq.~\eqref{eq:eikforwcont} is based on an upwind scheme which is expressed by \citep{treisterFastMarchingAlgorithm2016b}:
\begin{align}
  \label{eq:eikgodunov}
  & \max \left\{ - \sum_j D_{ij}^{+x} \: t_i, \: \sum_j D_{ij}^{-x} \: t_i, \: 0 \right\}^2   \nonumber \\
  &+  \max \left\{ - \sum_j D_{ij}^{+y} \: t_i, \: \sum_j D_{ij}^{-y} \: t_i, \:  0 \right\}^2 \nonumber \\
  &+  \max \left\{ - \sum_j D_{ij}^{+z} \: t_i, \: \sum_j D_{ij}^{-z} \: t_i, \: 0 \right\}^2  
  = \dfrac{1}{v_i^2} \, ,
\end{align}
where the superscript of the derivative matrices refers to the direction (forward or backward) and the coordinate axis (either $x$, $y$ or $z$) along which the derivative is taken.
The traveltime at a grid point (Fig.~\ref{fig:coarse_refined_grid}a) is then computed by solving a quadratic equation for $t_i$ resulting from  substituting the appropriate derivatives (either  eq.~\eqref{eq:fmmderiv2nd} or eq.~\eqref{eq:fmmderiv1st}) into eq.~\eqref{eq:eikgodunov} and choosing the largest root among the two solutions, since the computed traveltime must be greater than the other known traveltimes and in accordance with the entropy-satisfying solution \citep{sethian3DTraveltimeComputation1999,rawlinsonWaveFrontEvolution2004,treisterFastMarchingAlgorithm2016b}.
In our formulation, the position of the receivers, i.e., the location where measurements of the traveltimes are performed, is arbitrary and does not need to coincide with a grid point. So, once the traveltime at all grid points has been calculated, a bilinear (2D) or trilinear (3D) interpolation scheme is used to calculate the traveltime at the receiver positions. Other interpolation schemes could be implemented.

The major source of error in traveltime computations stems from the low accuracy of the predicted traveltime at the grid points around the source, because of the high curvature of the wavefront. One strategy to overcome such problem is to create a refined grid in the source region \citep{rawlinsonWaveFrontEvolution2004} so that the smaller grid spacing produces higher accuracy traveltimes and thus mitigates the issue. A 2D visual example is shown in Fig.~\ref{fig:coarse_refined_grid}, where in (b) the original grid is complemented with a refined grid around the source, with a grid spacing one third of that of the coarse grid. Implementation of such refinement of the grid has implications also for the inverse problem. Computation of gradients has to take into account the fine grid around the source.

\subsection{Setup of the inverse problem}

The final aim is to solve the inverse problem, i.e., to infer the velocity structure and source location from observed traveltimes. To do so, we first define an objective function, which, in this work, using matrix notation for a discretized problem, takes the following form
\begin{equation}
  \label{eq:objfunmatnot_to}
  \psi(\bs{u}(\bs{v}),\bs{t}_0,\bs{v}) = \dfrac{1}{2} \left( \bs{P} \, \bs{u}(\bs{v}) - (\bs{t}^p - \bs{t}^o) \right)^{\mr{T}} \bs{C}_{\mr{D}}^{-1} \left( \bs{P} \, \bs{u}(\bs{v}) - (\bs{t}^p - \bs{t}^o) \right) + \rho(\bs{v}) \, ,
\end{equation}
where $\bs{u}$ is the traveltime of the wavefront, $\bs{v}$ the seismic wave velocity, $\bs{t}^p$ are the picked arrival times, $\bs{t}^o$ the origin times, $\bs{P}$ a projection operator which takes the traveltime at each point on the grid ($\bs{u}$) and outputs the interpolated traveltime at the receiver' positions, $\bs{C}_{\mr{D}}$ a covariance matrix representing the uncertainty on the observed data, and, finally, $\rho(\bs{v})$ represents a prior term expressing our prior knowledge on the velocity model. From a probabilistic point of view \citep{mosegaard16ProbabilisticApproach2002,tarantolaInverseProblemTheory2005} eq.~\eqref{eq:objfunmatnot_to} represents the negative logarithm of the posterior probability density function (PDF), i.e., 
$ \sigma_{\rm M}(\bs{v}) = \mathit{const.} \: \exp(- \psi(\bs{u}(\bs{v}),\bs{v})) $.
While Eq.~\eqref{eq:objfunmatnot_to} is a particular choice of objective function, the theory developed here works for any objective function for which gradients can be calculated. As shown below, only the right-hand side of the adjoint equation will change upon changing the objective function.

The origin time $\bs{t}^o$ of a source is typically unknown and should be either estimated using independent information or inverted for. In the latter case, the derivative of the objective function with respect to the origin time is needed, unless gradient-free inversion methods are employed. This derivative is given by
\begin{equation}
  \label{eq:derivobjorigtime}
 \dfrac{\mr{d} \psi}{\mr{d} \bs{t}^o}  = \left( \bs{P} \, \bs{u} - (\bs{t}^p - \bs{t}^o) \right)^{\mr{T}} \bs{C}_{\mr{D}}^{-1} \, .
\end{equation}
For simplicity, and without loss of generality, we will ignore the origin time problem in the remainder of this paper. Hence, we define the observed data as $\bs{d} = \bs{t}^p - \bs{t}^o$ and rewrite the objective function as
\begin{equation}
  \label{eq:objfunmatnot}
  \psi(\bs{u}(\bs{v}),\bs{v}) = \dfrac{1}{2} \left( \bs{P} \, \bs{u}(\bs{v}) - \bs{d} \right)^{\mr{T}} \bs{C}_{\mr{D}}^{-1} \left( \bs{P} \, \bs{u}(\bs{v}) - \bs{d}  \right) + \rho(\bs{v}) \, .
\end{equation}

To solve the inverse problem, two strategies can be employed: i) the deterministic approach (optimization), which aims at minimizing eq.~\eqref{eq:objfunmatnot} to find an optimal model and ii) the probabilistic approach, which aims at characterising the posterior PDF and providing uncertainty quantification.
Strategy (i) typically requires the computations of the gradient of the objective function with respect to the model parameters (velocity and/or source location in this case). Strategy (ii), instead, typically solves non-linear inverse problems resorting to sampling algorithms. The latter methods in their original form \citep[e.g.,][]{metropolisEquationStateCalculations1953b,hastingsMonteCarloSampling1970} do not require gradients, however, the recently emerged Hamiltonian Monte Carlo (HMC) method \citep[e.g.,][]{duaneHybridMonteCarlo1987,girolamiRiemannManifoldLangevin2011,nealMCMCUsingHamiltonian2011,betancourtConceptualIntroductionHamiltonian2017}, which has been shown to work well on high-dimensional problems, does require computation of the gradients. This method has the advantage of exploiting information about the posterior PDF from gradients to drive the exploration of the model space by proposing transitions to distant states, reducing sample correlations, which has proved to be helpful in tomographic problems \citep[e.g.,][]{senTransdimensionalSeismicInversion2017,fichtnerHamiltonianMonteCarlo2019,aleardiHamiltonianMonteCarlo2020a,gebraadBayesianElasticFullWaveform2020,zuninoHamiltonianMonteCarlo2022a,zuninoHMCLabFrameworkSolving2023a,dhabariaHamiltonianMonteCarlo2024}.

In the following we explain how to efficiently compute gradients, which will be subsequently used for some numerical experiments using both deterministic and probabilistic strategies.

\section{Discrete adjoint method to compute gradients}
\label{sec:discradjgrad}

In this section we focus on obtaining gradients both for the velocity model (material parameters) and source location using the \emph{discrete} adjoint state method. First we describe the case of using a single grid and then the case of an additional refined grid around the source location.

We start by introducing the notation that will be used in the reminder of this work. The different sets of indices used hereafter are represented by $Q$, $P$ and $S$. The set $Q$ represents all the grid points.  The sets $S$ and $P$ are disjoint subsets of $Q$ for which $S \cup P = Q$, i.e., together they represent all grid points (both for the cases of fine and coarse grids). Specifically, the indices $q, k, j \in Q$ run on all grid node points. The indices $s, a \in S$ run on all grid points marked as source points in the grid, i.e., all the points belonging to the grid but where the traveltime has been computed using either an analytic formulation or a finer grid, while $p, i, b \in P$ run on all the grid points of the grid which have not been marked as source points. Finally, the indices $m$ and $n$ span the set of measurement points (observed data).

 In addition, we define two regions: region A, containing all points where the traveltime has been computed using either an analytic formulation or a finer grid (set $S$) and region B, which contains all remaining points (the traveltime in those points it is not calculated using the forward model $f_i$). See Fig.~\ref{fig:coarse_refined_grid} for a visual representation of the subdivision into regions and index notation for a 2D grid.

\subsection{Gradients with respect to velocity}

To compute the gradient of $\psi$ with respect to velocity, we first re-write the initial equations and the misfit functional in indicial notation. The implicit forward model in 3D then reads
\begin{equation}
  \label{eq:eikfwd3d}
  f_i = \Big( \sum_j D^x_{ij} {u}_j \Big)^2 + \Big( \sum_j D^y_{ij} u_j \Big)^2 + \Big( \sum_j D^z_{ij} {u}_j \Big)^2 - \dfrac{1}{v^{2}_i} = {0} \, ,
\end{equation}
where the index $i$ runs over all grid points in the B region (as the traveltimes in the source-defining A region are a fixed initial condition). The matrices $D^x_{ij}$, $D^y_{ij}$ and $D^z_{ij}$ represent the discrete spatial derivative operators, which contain the coefficients of the finite-difference stencils.
The misfit function is then given by
\begin{equation}
  \psi \left( \bs{u}(\bs{v}), \bs{v} \right) =
  \frac{1}{2}
  \sum_m 
  \Big( \Big( \sum_j P_{mj} u_j\Big) - d_m \Big)
  \, \sum_n  C^{-1}_{mn} \,
  \Big( \Big( \sum_j P_{nj} u_j\Big) - d_n \Big) + w(\bs{v})  \, ,
  \label{obj_func}
\end{equation}
while its total derivative with respect to velocity ($v$) is expressed by
\begin{equation}
  \label{eq:maingradvel}
  \dfrac{\mr{d} \psi}{\mr{d} v_q} =
   \sum_k \dfrac{\partial \psi}{\partial u_k} \dfrac{\mr{d} u_k}{\mr{d} v_q}   
   + \dfrac{\partial \psi}{\partial v_q} \, .
 \end{equation}
There are some dependencies to take into account to compute the gradient correctly. In particular, the traveltime $u$ is a function of different variables, depending on whether the point of interest is located in the region A or B. The dependencies are as follows
\begin{align}
  \label{eq:udepend_S}
  u_s &= u_s (v_{a \in S} ) \\ 
  \label{eq:udepend_P}
  u_p &= u_p \big( \underbrace{v_{b \in P}}_{\textrm{eikonal eq.}} ; \: \underbrace{ u_{s \in S} (v_{a \in S})}_{\textrm{initial cond.}} \big) \\
  \label{eq:psidepend_Q}
  \psi &= \psi \left( u_q (v_q), v_q ) \right) = \psi \left( u_p , u_s, v_q \right)
  \, .
\end{align}
In order to compute the gradient (a vector), we split the summation over $k$ in eq.~\eqref{eq:maingradvel} in two parts, the first related to the region A, i.e., spanning all the ``source'' points $s$ and the second spanning all points in the region B (see above). Moreover, we also subdivide the calculation of the gradient vector in two cases, one for $v_a$ and one for $v_b$. The split calculations read
\begin{equation}
  \label{eq:splitgradvel}
  \displaystyle \dfrac{\mr{d} \psi}{\mr{d} v_q} =
  \begin{cases}
    \displaystyle 
    \underbrace{ \sum_s  \dfrac{\partial \psi}{\partial u_s} \,  \dfrac{\mr{d} u_s}{\mr{d} v_a} }_{\textrm{T1Va}}
    +  \underbrace{\sum_p  \dfrac{\partial \psi}{\partial u_p} \dfrac{\mr{d} u_p}{\mr{d} v_a} }_{\textrm{T2Va}}
     \textrm{ for } q = a 
    \\[1.2cm]
    \displaystyle 
    \underbrace{ \sum_s  \dfrac{\partial \psi}{\partial u_s} \, \dfrac{\mr{d} u_s}{\mr{d} v_b} }_{\textrm{T1Vb } [=0]}
    +  \underbrace{\sum_p  \dfrac{\partial \psi}{\partial u_p} \dfrac{\mr{d} u_p}{\mr{d} v_b} }_{\textrm{T2Vb}} 
     \textrm{ for } q = b 
  \end{cases}
  + \underbrace{ \dfrac{\partial \psi}{\partial v_q}}_{\textrm{T3V}} \, ,
\end{equation}
where the term $\textrm{T1Vb}=0$ as $\frac{\mr{d} u_s}{\mr{d} v_b} = 0$ for $b \in P$, i.e., the derivative is equal to zero in the B region, as the traveltime inside the A region does not depend on the traveltime in the B region.
The derivative $\frac{\mr{d} \psi}{\mr{d} v_a}$ can thus be calculated by first computing individually the two terms T1Va and T2Va and adding them up, while for $\frac{\mr{d} \psi}{\mr{d} v_b}$ only one term, i.e., T2Vb needs to be calculated. Finally, the term T3V, i.e., $ \frac{\partial \psi}{\partial v_q}$ which usually represents the derivative of a prior term $\rho(\bs{m})$, is computed and added to the previous term to get the full gradient $\frac{\mr{d} \psi}{\mr{d} v_q}$. Below, we explain how to compute the three terms T1Va and T2Va and T2Vb, beginning with the latter for simplicity.

\subsubsection{Computing the term T2Vb for the gradient with respect to velocity}

The T2 terms of eq.~\eqref{eq:splitgradvel} involve a total derivative of wavefield $u$ with respect to velocity. Taking the derivative of the forward modelling operator, we find (see eq.~\eqref{eq:eikfwd3d}) 
\begin{equation}
  \dfrac{\mr{d} f_i}{\mr{d} v_b} = \sum_p \dfrac{\partial f_i}{\partial u_p} \dfrac{\mr{d} u_p}{\mr{d} v_b} + \dfrac{\partial f_i}{\partial v_b} = 0 \, ,
\end{equation}
after rearrangement, we get 
\begin{equation}
  \dfrac{\mr{d} u_p}{\mr{d} v_b} = - \sum_ i \left( \dfrac{\partial \bs{f}}{\partial \bs{u}} \right)_{\! pi}^{\!-1} \! \dfrac{\partial f_i}{\partial v_b} \, .
  \label{dupdvb}
\end{equation}
Substituting eq. \ref{dupdvb} into T2Vb (eq.~\eqref{eq:splitgradvel}), we find
\begin{equation}
  \label{eq:T2V_onb}
  \textrm{T2Vb} 
  = - \sum_p \sum_i \dfrac{\partial \psi}{\partial u_p}  \left( \dfrac{\partial \bs{f}}{\partial \bs{u}} \right)_{\! pi}^{\!-1} \!  \dfrac{\partial f_i}{\partial v_b}
  = \sum_i \lambda_i \dfrac{\partial f_i}{\partial v_b}
\end{equation}
where the variables $\lambda_i = - \sum_p \frac{\partial \psi}{\partial u_p}  \left( \frac{\partial f_i}{\partial u_p} \right)^{\!-1} $ are the adjoint state variables.
In order to obtain the term T2Vb we then follow a two-step procedure. First compute the adjoint variable $\lambda_i$ solving the following adjoint equation (a linear system)
\begin{equation}
  \label{eq:adjeq_velB}
  \sum_i \dfrac{\partial f_i}{\partial u_p} \lambda_i = - \dfrac{\partial \psi}{\partial u_p}
  \qquad  \left[ \mathrm{i.e.,} \,
    \left(\dfrac{\partial \bs{f}}{\partial \bs{u}} \right)^{\mr{T}}  \!\! \bs{\lambda} = - \left( \dfrac{\partial \psi}{\partial \bs{u}} \right)^{\mr{T}} \right] \, .
\end{equation}
The left-hand-side of the adjoint equation requires the derivative of eq.~\eqref{eq:eikfwd3d}
\begin{align}
  \label{eq:lhs_adj_df/du_k}
  \dfrac{\partial f_i}{\partial u_p} &= 2 \: \Big( \sum_j D^x_{ij} u_j \Big) D^x_{ip}
  + 2 \: \Big( \sum_j D^y_{ij} u_j \Big) D^y_{ip}  \nonumber \\
  & + 2 \: \Big( \sum_j D^z_{ij} u_j \Big) D^z_{ip}  \, ,
\end{align}
while the right-hand-side involves the derivative of eq.~\ref{obj_func}
\begin{equation}
  \label{eq:dpsi_dup}
  \dfrac{\partial \psi}{\partial u_p} =  \sum_m \Big[ \Big( \sum_j P_{mj} u_j\Big) - d_m \Big] \sum_n C^{-1}_{mn}  P_{np} \, .
\end{equation}
Note that $P_{np}$ and $D^x_{ip}$ have fewer columns than $P_{mj}$ and $D^x_{ij}$, as the derivative is over the index $p$ of $u_p$ (including only B region indices).

In matrix form, the adjoint equation can be written as
\begin{align}
  \label{eq:adjeqmatrixform}
  & \left[ 2 \: \mathrm{diag}\Big( \bs{D}^x \bs{u} \Big) \bs{D}^{xp}
    + 2 \: \mathrm{diag}\Big( \bs{D}^y \bs{u} \Big) \bs{D}^{yp} \right. \nonumber \\
  & \left. + 2 \: \mathrm{diag}\Big( \bs{D}^z \bs{u} \Big) \bs{D}^{zp} \right]^{\mr{T}} \bs{\lambda}
  = - \left[ \Big( \bs{P} \bs{u} - \bs{d} \Big)^{\mr{T}}  \bs{C}^{-1} \bs{P} \right]^{\mr{T}} \!\!\! ,
\end{align}
where $\bs{D}^x$, $\bs{D}^y$ and $\bs{D}^z$ represent the matrices containing the derivative coefficients in the $x$, $y$ and $z$ directions, and $\bs{D}^{xp}$, $\bs{D}^{yp}$ and $\bs{D}^{zp}$ are similar matrices where rows corresponding to positions in $A$ have been removed (those corresponding to the grid points around the source).

In addition to $\bs{\lambda}$, the T2Vb term (see eq.~\eqref{eq:T2V_onb}) requires also
\begin{equation}
  \dfrac{\partial f_i}{\partial v_b} = \dfrac{2}{v_{b}^3} \delta_{ib} \, ,
\end{equation}
i.e., a diagonal matrix with entries $\frac{2}{v^3_b}$, so that the T2Vb term becomes
\begin{equation}
 \mathrm{T2Vb} = \sum_p  \dfrac{\partial \psi}{\partial u_p} \dfrac{\mr{d} u_p}{\mr{d} v_b} =  \sum_i \lambda_i \: \dfrac{2}{v_b^3} \delta_{ib} = 2 \dfrac{\lambda_b}{v^3_b} 
\, .
\end{equation}

An important remark must be made regarding the practical solution of the adjoint equation \eqref{eq:adjeqmatrixform}. When using the FMM algorithm, in fact, the factor
\begin{align}
  \bs{A} &= \left[ 2 \: \mathrm{diag}\Big( \bs{D}^x \bs{u} \Big) \bs{D}^{xp} \right. \nonumber \\
  & \left. + 2 \: \mathrm{diag}\Big( \bs{D}^y \bs{u} \Big) \bs{D}^{yp}  + 2 \: \mathrm{diag}\Big( \bs{D}^z \bs{u} \Big) \bs{D}^{zp} \right] ,   
\end{align}
can be cast in a special form that renders the computation of gradients very efficient. The elements of the derivative matrices and traveltime vector can be arranged according to the increasing traveltime values encountered by the FMM algorithm while advancing the wavefront. In this way, the matrix $\bs{A}$ becomes lower triangular, which, upon transposition (see eq.~\eqref{eq:adjeqmatrixform}), turns into an upper triangular matrix, making the solution of the linear system for the adjoint equation much simpler and significantly more efficient than a generic case of an unstructured matrix. In addition, the $\bs{D}$ matrices are very sparse, since they only contain the coefficients of the derivative operators, which typically amount to 2-3 nonzero elements per row using a second-order FMM. Thus, exploiting FMM wavefront evolution ordering and sparsity of the matrices, special algorithms to solve the adjoint linear system can be used.

\subsubsection{Computing the term T2Va for the gradient with respect to\ velocity}

Traveltimes in the non-source region B ($u_p$) are also dependent on velocities in the source region A ($v_s$), via the traveltimes in source region A ($u_s(v_a)$).
We can thus write
\begin{align}
  \textrm{T2Va} &= \sum_p  \dfrac{\partial \psi}{\partial u_p} \dfrac{\mr{d} u_p}{\mr{d} v_a} \nonumber \\
  &= \sum_p \dfrac{\partial \psi}{\partial u_p}  \sum_s \dfrac{\partial u_p}{\partial u_s} \dfrac{\mr{d} u_s}{\mr{d} v_a} \, .
\end{align}
As $\frac{\partial u_p}{\partial u_s} =  \frac{\mr{d} u_p}{\mr{d} u_s}$, we exploit again the implicit derivation of the forward relation~\eqref{eq:eikfwd3d},
\begin{equation}
  \label{eq:dfi_dus}
  \frac{\mr{d} f_i}{\mr{d} u_s} = \frac{\partial f_i}{\partial u_p} \frac{\mr{d} u_p}{\mr{d} u_s} + \frac{\partial f_i}{\partial u_s}  = 0 \, ,
\end{equation}
to obtain
\begin{align}
  \label{eq:T2V_ona}
  \textrm{T2Va} &=  \sum_p \sum_s \dfrac{\partial \psi}{\partial u_p}  \dfrac{\partial u_p}{\partial u_s} \dfrac{\mr{d} u_s}{\mr{d} v_a} \nonumber \\
  &= - \sum_p \sum_i \sum_s \dfrac{\partial \psi}{\partial u_p} \left( \dfrac{\partial \bs{f}}{\partial \bs{u}} \right)_{\! pi}^{\!-1}  \! \! \dfrac{\partial f_i}{\partial u_s} \dfrac{\mr{d} u_s}{\mr{d} v_a}\nonumber  \\
  &=  \sum_i \sum_s \lambda_i \dfrac{\partial f_i}{\partial u_s} \dfrac{\mr{d} u_s}{\mr{d} v_a} \, ,
\end{align}
where $\lambda_i$ is the same adjoint variable as appears in eq.~\eqref{eq:T2V_onb}. Hence, the adjoint equation~\eqref{eq:adjeq_velB} needs to be solved only once to get $\lambda_i$ and then the solution can be used in both eqs.~\eqref{eq:T2V_onb} and \eqref{eq:T2V_ona}.

The remaining derivative that needs to be computed is that with respect to\ velocity on A, i.e., $\frac{\mr{d} u_s}{\mr{d} v_a}$. This term will be computed differently depending on whether there is a refinement of the grid near the source (see Fig.~\ref{fig:coarse_refined_grid}). If no refinement is applied, this can be found by differentiating equation~\eqref{eq:ttime_sourceregion}:
\begin{equation}
  \label{eq:dus_dva}
   \dfrac{\mr{d} u_s}{\mr{d} v_a} = -  \dfrac{ \sqrt{ \sum_r (n_{sr} - s_{r} )^2 } }{ \left( \sum_a \delta_{sa} v_a \right)^2} \, .
 \end{equation}
In the case of refinement of the grid in the source region, the above derivative is computed as explained in section~\ref{sec:gridrefin_vel}. The derivative with respect to the traveltime $u_s$ reads 
 \begin{align}
   \label{eq:parfi_parus}
   \dfrac{\partial f_i}{\partial u_s} &= 2 \:  \Big( \sum_j D^x_{ij} u_j \Big) D^x_{is} 
   + 2 \: \Big( \sum_j D^y_{ij} u_j \Big) D^y_{is}  \nonumber \\
   &+ 2 \: \Big( \sum_j D^z_{ij} u_j \Big) D^z_{is} \, ,
\end{align}
which allows us to finally compute the T2Va term as shown in eq.~\eqref{eq:T2V_ona}.

\subsubsection{Computing T1Va for the gradient with respect to\ velocity}

We now address the first term of the gradient T1Va (see eq.~\eqref{eq:splitgradvel}) involving a derivative with respect to\ to velocity in the A region. This term is:
\begin{equation}
  \label{eq:gradvelT1V}
  \mathrm{T1Va} = \sum_s  \dfrac{\partial \psi}{\partial u_s} \,  \dfrac{\mr{d} u_s}{\mr{d} v_a} \, .
\end{equation}
The first factor, i.e., $\frac{\partial \psi}{\partial u_s}$, takes the same form as eq.~\eqref{eq:dpsi_dup}, excepting the exchange of index $p$ with $s$:
\begin{equation}
  \label{eq:dpsi_dus}
  \dfrac{\partial \psi}{\partial u_s} =  \sum_m \sum_n \Big[ \Big( \sum_j P_{mj} u_j\Big) - d_m \Big] C^{-1}_{mn}  P_{ns} \, .
\end{equation}

The second factor, $\frac{d u_s}{d v_a}$ is computed differently depending on whether there is a refinement of the grid around the source or not. The next two sections address each of these cases: i) traveltimes $u_s$ calculated analytically, or ii) traveltimes $u_s$ computed using a refined grid around the source position as described in section~\ref{sec:fwdproblem}. 
In case there is no refinement of the grid around the source, and assuming the traveltime at the ``source'' points (A region) has been calculated using eq.~\eqref{eq:ttime_sourceregion}, the derivative of the traveltime at such grid nodes with respect to\ velocity is simply given by eq.~\eqref{eq:dus_dva}.
Computing $\frac{\mr{d} u_s}{\mr{d} v_a}$ in case of refinement of the grid near the source, instead, is explained in the next section (sec.~\ref{sec:gridrefin_vel}).

\subsubsection{Grid refinement around the source}
\label{sec:gridrefin_vel}

In the case of refinement of the grid around the source (i.e., region A), the procedure to compute the derivative of the traveltime in the coarse grid with respect to\ velocity at the same grid points, i.e., $\frac{\mr{d} u_s}{\mr{d} v_a}$, is similar to the one used in the coarse grid, as it also exploits the adjoint method. There are two   main differences: i) there is no misfit function, but the traveltime at coarse grid nodes, ii) there is an extra derivative term relating the velocity in the coarse grid to the interpolated velocity in the fine grid. 
For convenience and to avoid confusion, we slightly change notation for the traveltime in the source region (A region) computed on the coarse grid ($u_s$), by defining
\begin{equation}
  \hat{u}_h := u_s \, ,    
\end{equation}
where $\hat{u}_h$ represents the traveltime at points on the coarse grid in region A which are co-located with points in the fine grid and act as source points for the coarse grid.
The fine grid, in turn, contains a source region analogous to the case of the coarse grid without refinement. Therefore, for simplicity, we use again the same sets of indices for the fine grid as those devised before for the coarse grid. The variables in the \emph{refined} grid are labelled as $w$ for velocity and $\tau$ for traveltime.

We have the following dependencies inside the A region when grid refinement is present
\begin{align}
  \label{eq:wk_depend_Areg}
  w_q &= w_q ( v_{a \in A} ) \\
  \label{eq:tau_depend_Areg}
  \tau_k &= \tau_k ( w_{q \in A} (v_{a \in S})) \\
  \label{eq:us_depend_Areg}
    \hat{u}_h &= \hat{u}_h \left(\tau_{k \in A} \left(w_{q \in A} (v_{a \in S}) \right) \right) \, .
\end{align}
The velocity at the grid points inside the fine grid is interpolated as
\begin{equation}
  w_q = \sum_a N_{qa} v_a  \, ,
\end{equation}
where $N$ represents a matrix for linear interpolation, e.g., nearest neighbours. 
The derivative we aim to compute is $\frac{\mr{d} \hat{u}_h}{\mr{d} v_a} $, with total derivative
\begin{align}
  \label{eq:gradvel_finegrid}
  \dfrac{\mr{d} \hat{u}_h}{\mr{d} v_a}
  &= \sum_k \dfrac{\partial \hat{u}_h}{\partial \tau_k} \dfrac{\mr{d} \tau_k}{\mr{d} v_a} + \underbrace{ \dfrac{\partial \hat{u}_h}{\partial v_a} }_{=0} \nonumber \\
  &= \sum_q \sum_k \dfrac{\partial \hat{u}_h}{\partial \tau_k } \left( \dfrac{\partial \tau_k}{\partial w_q} \dfrac{\mr{d} w_q}{\mr{d} v_a} + \underbrace{ \dfrac{\partial \tau_k}{\partial v_a} }_{=0} \right) \nonumber \\
  &= \sum_q \sum_k \dfrac{\partial \hat{u}_h}{\partial \tau_k } \dfrac{\partial \tau_k}{\partial w_q} \dfrac{\mr{d} w_q}{\mr{d} v_a} 
\end{align}
where $\frac{\partial \tau_k}{\partial v_a} = 0$ and $\frac{\partial \hat{u}_h}{\partial v_a} = 0$ because $\hat{u}_h$ and $\tau_k$ do not depend directly on $v_a$, see eqs.~\eqref{eq:tau_depend_Areg} and \eqref{eq:us_depend_Areg}.
  We first focus on the computation of $ \frac{\partial \hat{u}_h}{\partial \tau_k} \frac{\partial \tau_k}{\partial w_q} $, which can be partly computed by means of the adjoint method as described in the following. Since $\frac{\mr{d} \tau_k}{\mr{d} w_q} = \frac{\partial \tau_k}{\partial w_q}$ (see eq.~\eqref{eq:tau_depend_Areg}), we have
\begin{equation}
\frac{\mr{d} \hat{u}_h}{\mr{d} w_q} = \sum_k \frac{\partial \hat{u}_h}{\partial \tau_k} \frac{\mr{d} \tau_k}{\mr{d} w_q} \, .    
\end{equation}
In analogy with the solution illustrated for the coarse grid (see eq.~\eqref{eq:splitgradvel}), we split the derivative into four terms for calculation, in analogy to eq.~(\ref{eq:splitgradvel}). We thus obtain
\begin{equation}
  \label{eq:splitgradvelrefin}
  \displaystyle \dfrac{\mr{d} \hat{u}_h}{\mr{d} w_q} =
  \begin{cases}
    \displaystyle 
    \underbrace{ \sum_s  \dfrac{\partial \hat{u}_h}{\partial \tau_s} \,  \dfrac{\mr{d} \tau_s}{\mr{d} w_a} }_{\textrm{T1Wa}}
    +  \underbrace{\sum_p  \dfrac{\partial \hat{u}_h}{\partial \tau_p} \dfrac{\mr{d} \tau_p}{\mr{d} w_a} }_{\textrm{T2Wa}}
     \textrm{ for } q = a 
    \\[1.2cm]
    \displaystyle 
    \underbrace{ \sum_s  \dfrac{\partial \hat{u}_h}{\partial \tau_s} \, \dfrac{\mr{d} \tau_s}{\mr{d} w_b} }_{\textrm{T1Wb } [=0]}
    +  \underbrace{\sum_p  \dfrac{\partial \hat{u}_h}{\partial \tau_p} \dfrac{\mr{d} \tau_p}{\mr{d} w_b} }_{\textrm{T2Wb}} 
     \textrm{ for } q = b 
  \end{cases} \, ,
\end{equation}
where the term $\textrm{T1Wb}=0$ as $\frac{\partial \tau_s}{\partial w_b} = 0$, i.e., this derivative is equal to zero in the B region. The derivative $\frac{\mr{d} \hat{u}_h}{\mr{d} w_a}$ can thus be calculated by first computing individually the two terms T1Wa and T2Wa and adding them, while for $\frac{\mr{d} \hat{u}_h}{\mr{d} w_b}$ only one term, i.e., T2Wb needs to be calculated. 

The equation analogous to the implicit forward modeling expression for the fine grid reads
\begin{equation}
  \label{eq:fwd_finegrid}
  g_i = \Big( \sum_j D^x_{ij} \tau_j \Big)^2  + \Big( \sum_j D^y_{ij} \tau_j \Big)^2 + \Big( \sum_j D^z_{ij} \tau_j \Big)^2 - \dfrac{1}{w^{2}_i} = {0} \, .
\end{equation}
As we invoke the fine grid to determine $\frac{d \hat{u}_h}{d v_s}$, the traveltime at the source points for the coarse grid now takes the role of the objective function in the previous calculation. We write this as 
\begin{equation}
\hat{u}_h = \sum_q H_{hq} \tau_q \, ,
\end{equation}
where $H_{hq}$ is a sampling operator extracting the traveltime from the fine grid at the grid points co-located with the coarse grid, represented by the index $h$ (see Fig.~\ref{fig:coarse_refined_grid}). 
Since, in contrast to the objective function $\psi$, the traveltime $\hat{u}_h$ is not a scalar, the adjoint variable is as a matrix instead of a vector, as shown in the following.
In strict analogy to the coarse grid derivation of T2Vb (see eq.~\ref{eq:eikfwd3d} - \ref{eq:adjeq_velB}), the expression for T2Wb is 
\begin{align}
  \label{eq:T2Wb_adj}
  \mathrm{T2Wb} =  \sum_i \Lambda_{hi} \dfrac{\partial g_i}{\partial w_q},
\end{align}
where the adjoint source matrix $\bs{\Lambda}$ satisfies
\begin{equation}
  \label{eq:adjeq_finegrdV}
  \sum_i \dfrac{\partial g_i}{\partial \tau_p} \Lambda_{ih} = - \dfrac{\partial \hat{u}_h}{\partial \tau_p} \qquad  \left[ \mathrm{i.e.,} \, \left(\dfrac{\partial \bs{g}}{\partial \bs{\tau}} \right)^{\mr{T}}  \!\! \bs{\Lambda} = - \left( \dfrac{\partial \bs{u}}{\partial \bs{\tau}} \right)^{\mr{T}} \right] \, .
\end{equation}
The right-hand side of the adjoint equation~\eqref{eq:adjeq_finegrdV} is calculated as
\begin{equation}
  \label{eq:du_h_dtau_p}
  \dfrac{\mr{d} \hat{u}_h}{\mr{d} \tau_p} = H_{hp} \, ,
\end{equation}
while the left-hand side\ is obtained in the same way as eq.~\eqref{eq:lhs_adj_df/du_k}. We arrive at the final expression for T2Wb
\begin{align}
   \mathrm{T2Wb} =  2 \dfrac{\Lambda_{hb}}{w_b^3}  \, .
\end{align}

The next term to be addressed is
\begin{align}\mathrm{T2Wa} &= \sum_p  \dfrac{\partial \hat{u}_h}{\partial \tau_p} \dfrac{\mr{d} \tau_p}{\mr{d} w_a}
= \sum_i \sum_s \Lambda_{hi} \dfrac{\partial g_i}{\partial \tau_s} \dfrac{\mr{d} \tau_s}{\mr{d} w_a}, \, 
\end{align}
where $\Lambda_{hi}$ is the same as in eq.~\eqref{eq:T2Wb_adj}, i.e., it solves eq.~\eqref{eq:adjeq_finegrdV}.
In analogy with eq.~\eqref{eq:parfi_parus} the remaining derivative is computed as 
 \begin{equation}
   \label{eq:pargi_partaus}
  \dfrac{\partial g_i}{\partial \tau_s} = 2 \:  \Big( \sum_j D^x_{ij} \tau_j \Big) D^x_{is} + 2 \: \Big( \sum_j D^y_{ij} \tau_j \Big) D^y_{is} + 2 \: \Big( \sum_j D^z_{ij} \tau_j \Big) D^z_{is} \, .
\end{equation}

We now focus on the T1Wa term, i.e., 
\begin{equation}
  \mathrm{T1Wa} = \sum_s  \dfrac{\partial \hat{u}_h}{\partial \tau_s} \, \dfrac{\mr{d} \tau_s}{\mr{d} w_a} \, .
\end{equation}
The first factor is calculated as 
\begin{equation}
  \label{eq:du_h_dtau_s}
  \dfrac{\mr{d} \hat{u}_h}{\mr{d} \tau_s} = H_{hs} \, ,
\end{equation}
while, in analogy to eq.~\eqref{eq:dus_dva}, the second factor is simply
\begin{equation}
  \label{eq:dtaus_dwa}
  \dfrac{\mr{d} \tau_s}{\mr{d} w_a} =  -  \dfrac{\sqrt{ \sum_r (x_{sr}^{\textrm{fine grid}} - s_{sr}^{\textrm{fine grid}} )^2 } }{ \left( \sum_a \delta_{sa} v_a \right)^2} \, .
\end{equation}

The last factor that remains to be computed from eq.~\eqref{eq:gradvel_finegrid} is $\frac{\mr{d} w_q}{\mr{d} v_a}$. The derivative reads
\begin{equation}
  \frac{\mr{d} w_q}{\mr{d} v_a}  = N_{qa} \,.
\end{equation}

\subsection{Gradient with respect to source location}

We now focus on the computation of the gradient of the misfit functional  with respect to\ the source location. This gradient is calculated in two steps, by means of the chain rule. The first step concerns the derivative of the misfit with respect to\ traveltime at the grid nodes outside the source region (B region), while the second step relates to the derivative of traveltime at such source grid points with respect to the position of the source (A region). The misfit functional can be expressed as below, together with the following dependencies:
\begin{align}
   u_s &= u_s(s_r) \\
   u_p &= u_p(u_s(s_r)) \\
   \psi &= \psi(u_q(s_r),s_r ) = \psi(u_p(u_s(s_r)),u_s (s_r),s_r )  \, ,
\end{align}
where $s_r$ represent the coordinates of the source and $r=1,2,3$ in 3D, i.e., the three coordinates $(x,y,z)$ of a given source. The total derivative of the objective function with respect to source location then reads
\begin{align}
  \label{eq:gradsrcnoref}
  \dfrac{\mr{d} \psi}{\mr{d} s_r}
  &= \sum_q \dfrac{\partial \psi}{\partial u_q} \dfrac{\mr{d} u_q}{\mr{d} s_r} + \dfrac{\partial \psi}{\partial s_r} \\
  \label{eq:dpsi_dup_srcloc}
  &= \sum_p \sum_s \dfrac{\partial \psi}{\partial u_p} \dfrac{\mr{d} u_p}{\mr{d} u_s} \dfrac{\mr{d} u_s}{\mr{d} s_r} + \sum_s \dfrac{\partial \psi}{\partial u_s} \dfrac{\mr{d} u_s}{\mr{d} s_r} + \dfrac{\partial \psi}{\partial s_r} \, ,
 \end{align}
where we split the summation over $q$ in two terms and applied the chain rule for $u_p$. 
To avoid computing the factor $\frac{\mr{d} u_p}{\mr{d} u_s}$ we resort, once again, to the adjoint method, extracting the derivative from the implicit forward model (eq.~\eqref{eq:eikfwd3d}), to obtain
\begin{equation}
  \label{eq:dup_dus_srcloc}
  \dfrac{\mr{d} u_p}{\mr{d} u_s} = - \sum_i \left( \dfrac{\partial \bs{f}}{\partial \bs{u}}  \right)_{\! pi}^{\! -1} \! \dfrac{\partial f_i}{\partial u_s} \, ,
\end{equation}
which, once substituted in the expression for the gradient, gives
\begin{align}
  \dfrac{\mr{d} \psi}{\mr{d} s_r}
  &= - \sum_p \sum_i \dfrac{\partial \psi}{\partial u_p}
   \left( \dfrac{\partial \bs{f}}{\partial \bs{u}}  \right)_{\! pi}^{\! -1} \! \dfrac{\partial f_i}{\partial u_s}
    \dfrac{\mr{d} u_s}{\mr{d} s_r} + \sum_s \dfrac{\partial \psi}{\partial u_s} \dfrac{\mr{d} u_s}{\mr{d} s_r} + \dfrac{\partial \psi}{\partial s_r} \nonumber \\
  \label{eq:dup_dus_adj_srcloc}
  &= \sum_p \sum_i \lambda_i \dfrac{\partial f_i}{\partial u_s}
    \dfrac{\mr{d} u_s}{\mr{d} s_r} + \sum_s \dfrac{\partial \psi}{\partial u_s} \dfrac{\mr{d} u_s}{\mr{d} s_r} + \dfrac{\partial \psi}{\partial s_r} \, ,
\end{align}
where the adjoint variable $\bs{\lambda}$ is once again that calculated in equation~\eqref{eq:adjeq_velB}.
The only remaining factor to compute to get the first term for the derivative $\frac{\mr{d} \psi}{\mr{d} s_r}$ is $\frac{\partial f_i}{\partial u_s}$, which is computed as shown in eq.~\eqref{eq:parfi_parus}.

As the traveltime from the source location to the enclosing grid point is computed analytically using eq.~\eqref{eq:ttime_sourceregion}, the derivative $\frac{\mr{d} u_s}{\mr{d} s_r}$ is given by
\begin{equation}
  \label{eq:gradussrcpos}
  \dfrac{\mr{d} u_s}{\mr{d} s_r} = \dfrac{ - (n_{sr} - s_r)}{ v_s \sqrt{ \sum_r (n_{sr} - s_r )^2 } } \, .
\end{equation}

Finally $\frac{\partial \psi}{\partial u_s}$, can be calculated using eq.~\eqref{eq:dpsi_dus}.
Combining all the above, we can now compute the gradient with respect to the source position as expressed by eq.~\eqref{eq:gradsrcnoref}.

\subsubsection{Gradient with respect to source location with grid refinement}

In the case of refinement of the grid around the source, the gradient with respect to the source location is computed as follows. First, we explicitly state the dependencies for the variables involved in this setup
\begin{align}
  u_s &= u_s(\tau_s(s_r)) \\
  u_p &= u_p(u_s(\tau_p(s_r))) \\
  \psi &= \psi(u_q(s_r),s_r ) = \psi(u_p(u_s(\tau_q(s_r))),u_s(\tau_s (s_r)),s_r )  \, .
\end{align}
The gradient is computed as in eq.~\eqref{eq:dup_dus_adj_srcloc} except for the factor $\frac{\mr{d} u_s}{\mr{d} s_r} = \frac{\mr{d} \hat{u}_h}{\mr{d} s_r}$, which, in this case, depends on the values obtained in the fine grid. We have
\begin{align}
  \label{eq:du_h_ds_r_adj_srcloc}
  \dfrac{\mr{d} \hat{u}_h}{\mr{d} s_r}
  &= \sum_q \dfrac{\partial \hat{u}_h}{\partial \tau_q} \dfrac{\mr{d} \tau_q}{\mr{d} s_r} + \dfrac{\partial \hat{u}_h}{\partial s_r} \nonumber \\
  &= \sum_p \sum_s \dfrac{\partial \hat{u}_h}{\partial \tau_p} \dfrac{\mr{d} \tau_p}{\mr{d} \tau_s}  \dfrac{\mr{d} \tau_s}{\mr{d} s_r} + \sum_s  \dfrac{\partial \hat{u}_h}{\partial \tau_s} \dfrac{\mr{d} \tau_s}{\mr{d} s_r} + \dfrac{\partial \hat{u}_h}{\partial s_r} \nonumber \\
  &= \sum_p \sum_i \sum_s \Lambda_{hi} \dfrac{\partial g_i}{\partial \tau_s} \dfrac{\mr{d} \tau_s}{\mr{d} s_r} + \sum_s \dfrac{\partial \hat{u}_h}{\partial \tau_s} \dfrac{\mr{d} \tau_s}{\mr{d} s_r} + \dfrac{\partial \hat{u}_h}{\partial s_r} 
  \, .
\end{align}
The above relation is analogous to eq.~\eqref{eq:dpsi_dup_srcloc} and eq.~\eqref{eq:dup_dus_adj_srcloc}. Here, again $\Lambda_{ih}$ is the adjoint variable which provides the solution to eq.~\eqref{eq:adjeq_finegrdV}.

In summary, to get the derivative $\frac{\mr{d} u_s}{\mr{d} s_r} = \frac{\mr{d} \hat{u}_h}{\mr{d} s_r}$ the following calculations have to be performed:
\begin{enumerate}
\item Solve the adjoint equation~\eqref{eq:adjeq_finegrdV}, where the left-hand side is obtained analogously to  eq.~\eqref{eq:lhs_adj_df/du_k}, using the variables of the refined grid instead of the those for the coarse grid and the r.h.s.\ by eq.~\eqref{eq:du_h_dtau_p};
\item Compute $\frac{\partial g_i}{\partial \tau_s}$ as in eq.~\eqref{eq:pargi_partaus};
\item Compute $\frac{\mr{d} \tau_s}{\mr{d} s_r}$ as in eq.~\eqref{eq:gradussrcpos};
\item Compute $\frac{\partial \hat{u}_h}{\partial \tau_s}$ as in eq.~\eqref{eq:du_h_dtau_s};
\item Combine all derivatives as shown in eq.~\eqref{eq:du_h_ds_r_adj_srcloc}.
\end{enumerate}

\section{Numerical experiments}
\label{sec:numexper}

In the following we present a set of numerical experiments where the gradients calculated using the formalism exemplified above is employed for both deterministic and probabilistic inversion examples.
In the first case, we use an L-BFGS algorithm \citep[e.g.,][]{nocedalUpdatingQuasiNewtonMatrices1980,nocedalNumericalOptimization2006}, a limited-memory quasi-Newton algorithm which relies on an approximation to the inverse Hessian matrix.
In the second case, we employ the HMC method \citep[e.g.,][]{nealMCMCUsingHamiltonian2011,fichtnerHamiltonianMonteCarlo2019}, which is based on a sampling algorithm that exploits the gradient of the posterior PDF to address high-dimensional problems. We refer the reader to \citet{fichtnerHamiltonianMonteCarlo2019} and \citet{zuninoHMCLabFrameworkSolving2023a} for details of the method and its implementation.

The codes used to implement forward and gradient computations (``EikonalSolvers.jl'') and to perform the deterministic and probabilistic inversions (``InverseAlgos.jl'') are written in the Julia language \citep{bezansonJuliaFreshApproach2017} and are part of the open-source framework G$^{-1}$Lab/HMCLab \citep{zuninoHMCLabFrameworkSolving2023a}, currently maintained at the following address \url{https://github.com/GinvLab}. 

\subsection{Optimization examples}

\begin{figure}
    \centering
  \includegraphics[width=\textwidth]{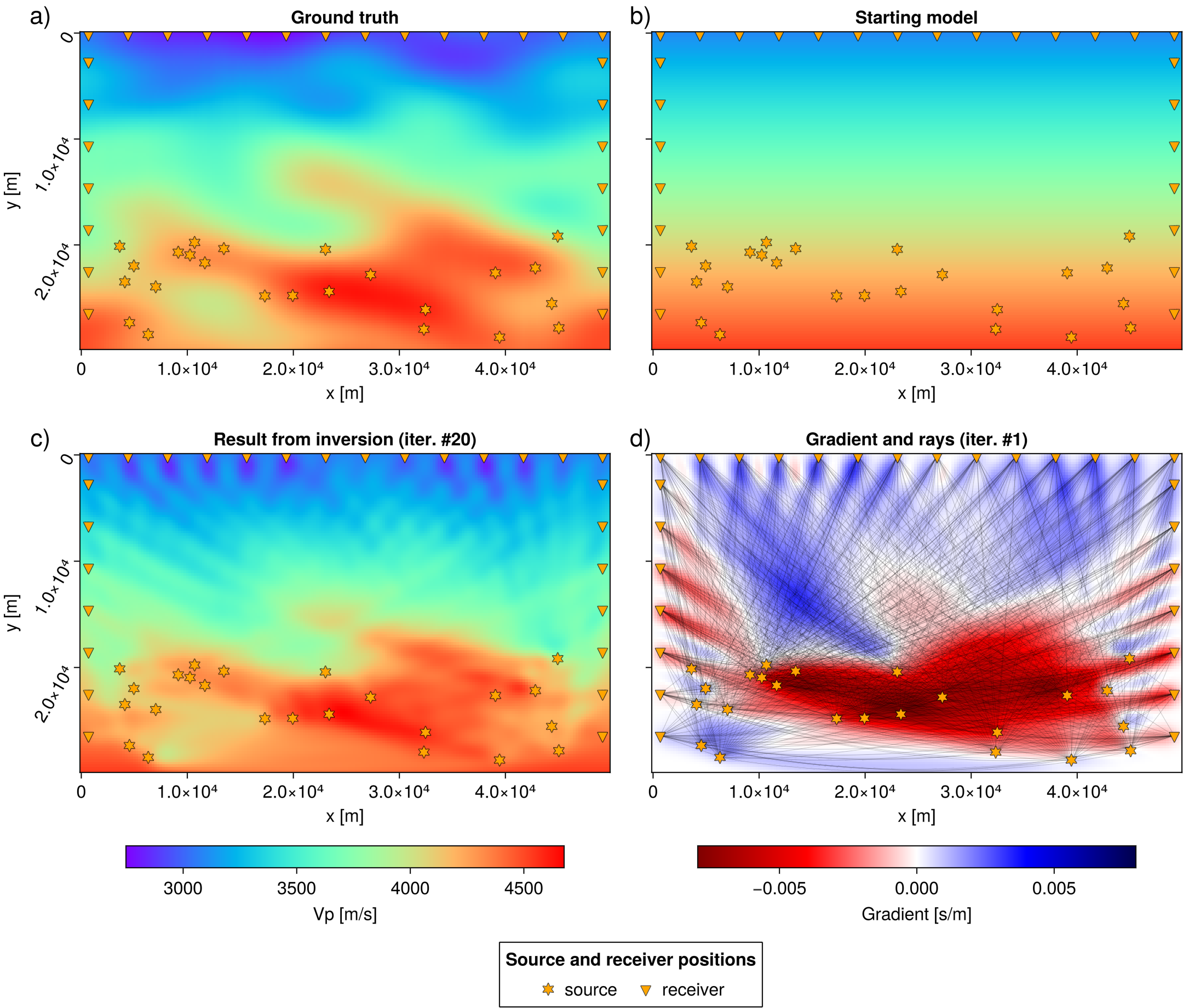}
  \caption{Deterministic inversion for a 2D velocity model. a) Ground-truth model from which the observed data (with the addition of noise) have been generated. b) Starting model, a laterally-homogeneous velocity model. c) The result from inversion after 20 iterations of the L-BFGS algorithm. d) An example of the gradient with respect to velocity at the first iteration with rays overlaid for comparison. Source and receiver positions are shown as stars and triangles, respectively.}
  \label{fig:bfgs_vel2D}
\end{figure}
The first example we present is an inversion targeting only velocity (known source locations). The geometry of the problem features a 2D model consisting of 200$\times$120 grid points for a total of 24000 model parameters spanning an area of about 3$\times$5 km, with a grid spacing of 250 m. The 28 receivers are located on top and on the left and right sides of the model, while the 24 sources are randomly distributed in the lower part of the model (see Fig.~\ref{fig:bfgs_vel2D}). The artificial data are generated from a synthetic model (Fig.~\ref{fig:bfgs_vel2D}a) with the addition of Gaussian noise to simulate real data. We employ a simple diagonal covariance matrix to describe the uncertainty of the traveltime data. The starting model (Fig.~\ref{fig:bfgs_vel2D}) for the inversion is a laterally homogeneous velocity model, and, for simplicity, we assume an unbounded uniform prior.
An example of the gradient obtained in the first iteration is shown in Fig.~\ref{fig:bfgs_vel2D}d, along with ray paths for comparison. To solve the optimisation problem, we run 20 iterations of the L-BFGS algorithm, i.e., until the change in misfit becomes negligible, and show the solution in Fig.~\ref{fig:bfgs_vel2D}c. The result appears fairly close the ground truth, apart from some imprint of the source-receiver geometry particularly in the shallow part of the model and some less resolved areas in the centre-left bottom part. 

\begin{figure}
    \centering
  \includegraphics[width=\textwidth]{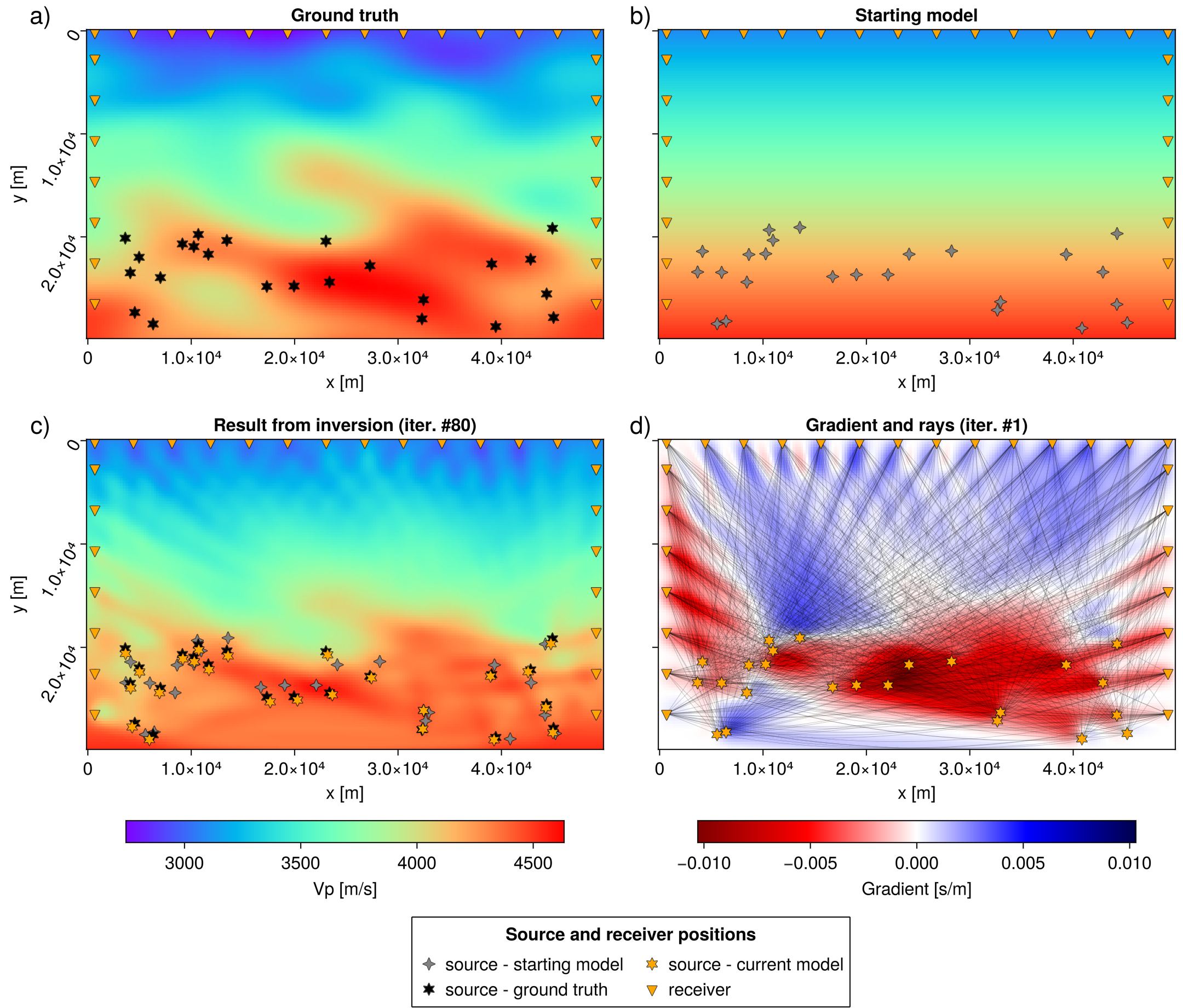}
  \caption{Deterministic inversion for both a 2D velocity model and source locations. a) Ground-truth model (velocity and source locations) from which the artificial data (with the addition of noise) have been generated. b) Starting model, a laterally-homogeneous velocity model and location of sources produced by randomly perturbing the ground-truth positions to within a maximum distance. c) The result from inversion after 150 iterations of the L-BFGS algorithm. d) An example of the gradient with respect to velocity in the first iteration with rays overlaid for comparison. Source locations are shown as different kinds of stars (see the legend) and receiver positions are shown as upside-down triangles.}
  \label{fig:bfgs_velsrcloc2D}
\end{figure}
The second example features the same geometry and parameters as the first one, with the addition of inverting for both the velocity model and source locations. In this case, in order to run the L-BFGS algorithm, we thus compute the gradient of the objective function both with respect to velocity and source location. The ground truth and starting models for velocity (see Fig.~\ref{fig:bfgs_velsrcloc2D}a and \ref{fig:bfgs_velsrcloc2D}b are the same as for the first example. The starting model for source locations is a random perturbation of the ground truth positions with a maximum radius of 1500 m.  An example of the gradient obtained at the first iteration is shown in Fig.~\ref{fig:bfgs_velsrcloc2D}d, along with ray paths for comparison.
The L-BFGS algorithm is run until the change in the misfit function is deemed negligible, which results in 80 iterations. The final result (see Fig.~\ref{fig:bfgs_velsrcloc2D}c) shows a good agreement with the ground truth model both in terms of velocity and source positions. As in the previous case, some imprint of the source-receiver paths is visible in the velocity model, and some areas, particularly at the bottom, appear less resolved than the inner part of the model. The reconstructed position of the sources, as expected, is not perfect, although the distance to the ground truth locations is on average fairly small. The main difference in comparison with the first numerical experiment is that the number of L-BFGS iterations required to obtain a similar reconstruction in terms of velocity is much larger. This is to be expected, since the inverse problem for the case where  the source positions are unknown is a significantly more difficult problem to solve. There is, in fact, a trade-off between the two parameters and, therefore, the L-BFGS needs more iterations to optimize the values for both the velocity model and the source locations.

\begin{figure}
    \centering
  \includegraphics[width=\textwidth]{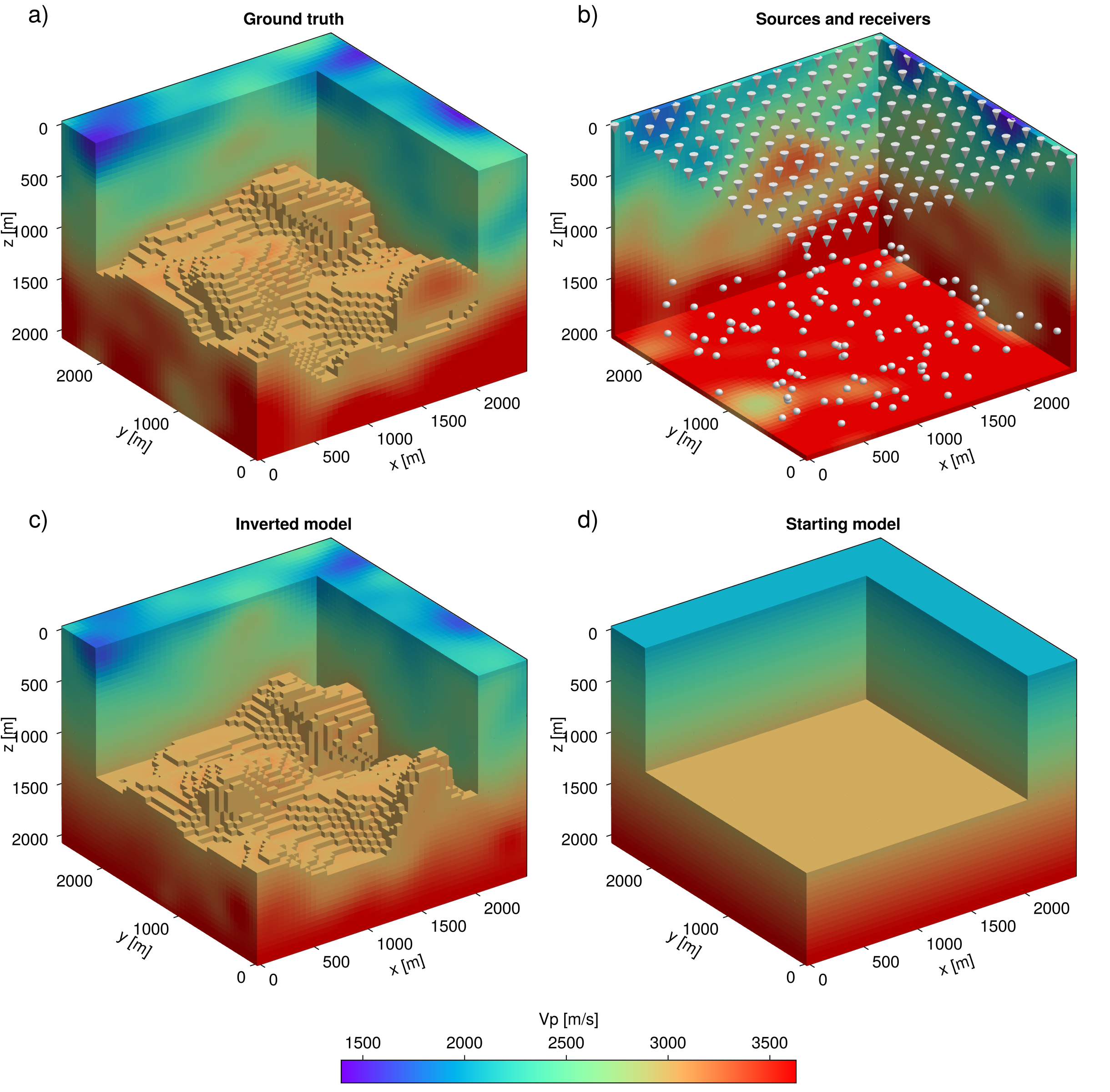}
  \caption{Deterministic inversion for a 3D velocity model. a) Ground-truth model from which the observed data (with the addition of noise) have been generated. b) Source (spheres) and receiver (cone) locations. c) The result from inversion after 40 iterations of the L-BFGS algorithm. d) Starting model, a laterally-homogeneous velocity model and location of sources.}
  \label{fig:bfgs_vel3D}
\end{figure}
The third example is a 3D problem targeting only the velocity structure of a synthetic model (Fig.~\ref{fig:bfgs_vel3D}a), where the sources are scattered in the bottom part of the model (Fig.~\ref{fig:bfgs_vel3D}b) and all receivers are at the surface. Figure~\ref{fig:bfgs_vel3D} shows the ground truth model (a) and the geometry of sources and receivers (b). The inversion has been carried out again using the l-BFGS algorithm for 50 iterations, again stopping when the change in the misfit was negligible. The starting model is laterally homogeneous (Fig.~\ref{fig:bfgs_vel3D}d), with velocity linearly increasing with depth. The final model resulting from the inversion is shown in Fig.~\ref{fig:bfgs_vel3D}c, where a good agreement with the ground truth model is found. In particular, the recovered pattern of the subsurface structure is close to the ground-truth model.

\subsection{Hamiltonian Monte Carlo examples}

\begin{figure}
    \centering
  \includegraphics[width=\textwidth]{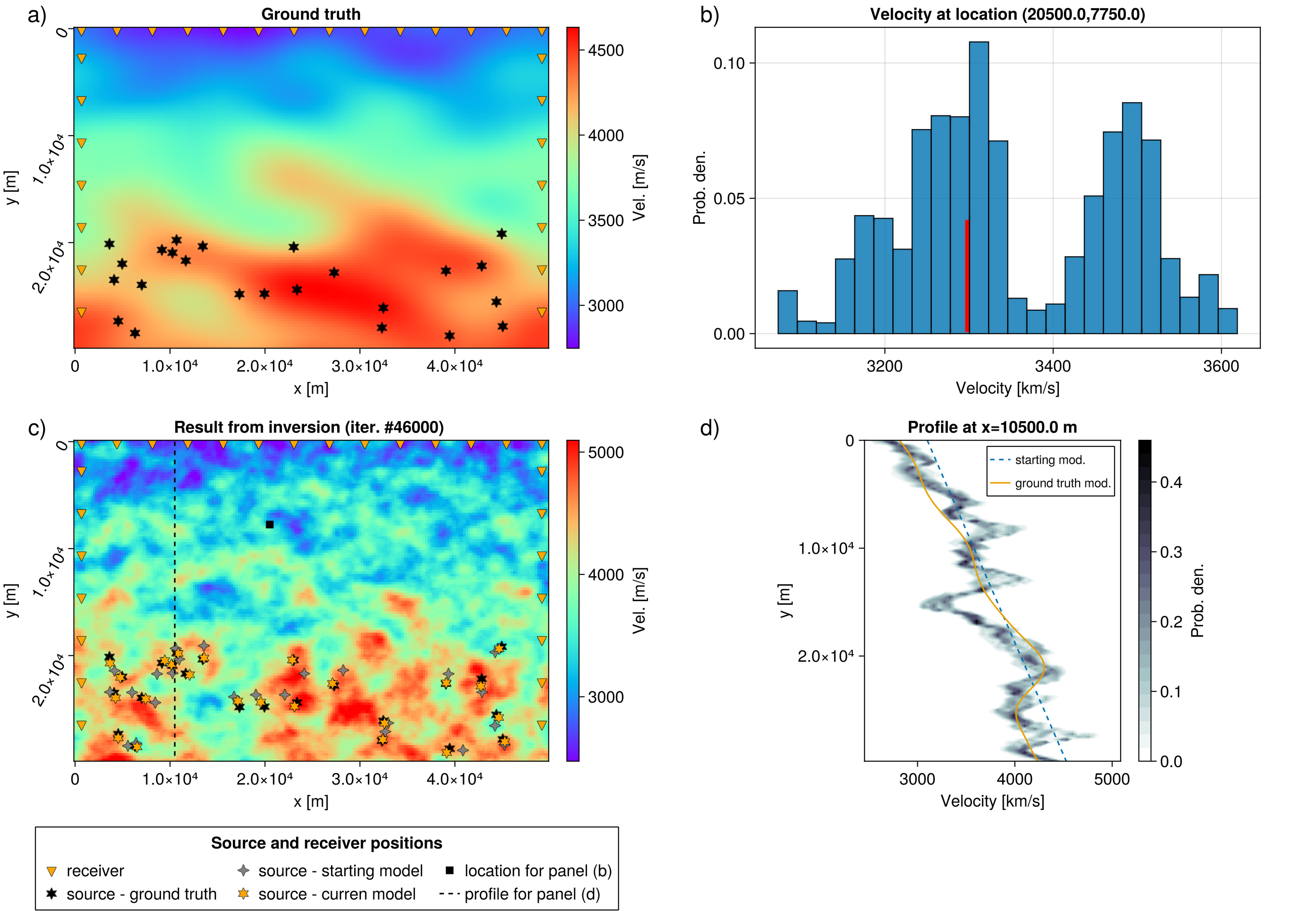}
  \caption{Probabilistic inversion for both a 2D velocity model and source locations. a) Ground-truth model (velocity and source locations) from which the observed data (with the addition of noise) have been generated. b) Histogram of velocity at a selected location, marked by the black square in (c). c) A random model from the posterior collection after 46000 iterations. d) A vertical profile at x=10500 m showing the starting model, the ground truth model and the PDF computed from the collection of posterior models.}
  \label{fig:hmc_velsrcloc2D}
\end{figure}
In analogy with the previous optimization examples, we have set up two synthetic experiments with a probabilistic approach. In particular, the sampling algorithm employed is a flavour of the HMC method \citep{duaneHybridMonteCarlo1987,nealMCMCUsingHamiltonian2011}, the No U-Turn (NUTS) algorithm \citep{hoffmanNoUturnSamplerAdaptively2014}, which features an adaptive number of integration steps of the Hamiltonian dynamics to auto-tune some of the algorithm's parameters.

The first example shown in the following features the same geometry and ground-truth model than those used for the optimization tests. The target of the inversion is both the velocity structure and the source location. The NUTS algorithm has been run for 50000 iterations, starting from a laterally-homogeneous model as shown in Figure~\ref{fig:bfgs_velsrcloc2D}b. After burn-in, the collection of remaining models (150000 were saved) represents a set of samples of the posterior PDF.  Figure~\ref{fig:hmc_velsrcloc2D}a shows the ground-truth model and the position of the sources. A histogram of velocity values at a particular location (see Fig.~\ref{fig:hmc_velsrcloc2D}c) is depicted in Fig.~\ref{fig:hmc_velsrcloc2D}b, demonstrating multi-modality in the posterior PDF for certain parameters. In this particular case, there are two ranges of velocity which sport a higher probability than other values, centred around 3500 and 3800 m/s, respectively. This example illustrates the usefulness of performing uncertainty quantification using probabilistic methods, as, if a deterministic method was used, it would not have been possible to detect such feature. An example of a model of both velocity structure and the source locations from the posterior collection is shown in Fig.~\ref{fig:hmc_velsrcloc2D}c. As expected, the source locations of the posterior model are closer to the ground truth than the initial model, as is the velocity structure. The velocity model shares some long-wavelength features with the ground-truth model, however, the more detailed pattern appears to be more random, as is typical for a single model drawn from a probabilistic approach. The last panel (Fig.~\ref{fig:hmc_velsrcloc2D}d) depicts an example vertical profile of velocity located at $x=10500$ m (marked as a dashed line in Fig.~\ref{fig:hmc_velsrcloc2D}c), where the a contouring of the resulting PDF, i.e., a marginal, is compared to the starting and ground-truth model. The contouring obtained from the posterior collection of models partly follows the ground-truth model, however it deviates particularly around 16500 m, where the inversion results have likely been trapped in a local minimum.
\begin{figure}
    \centering
  \includegraphics[width=\textwidth]{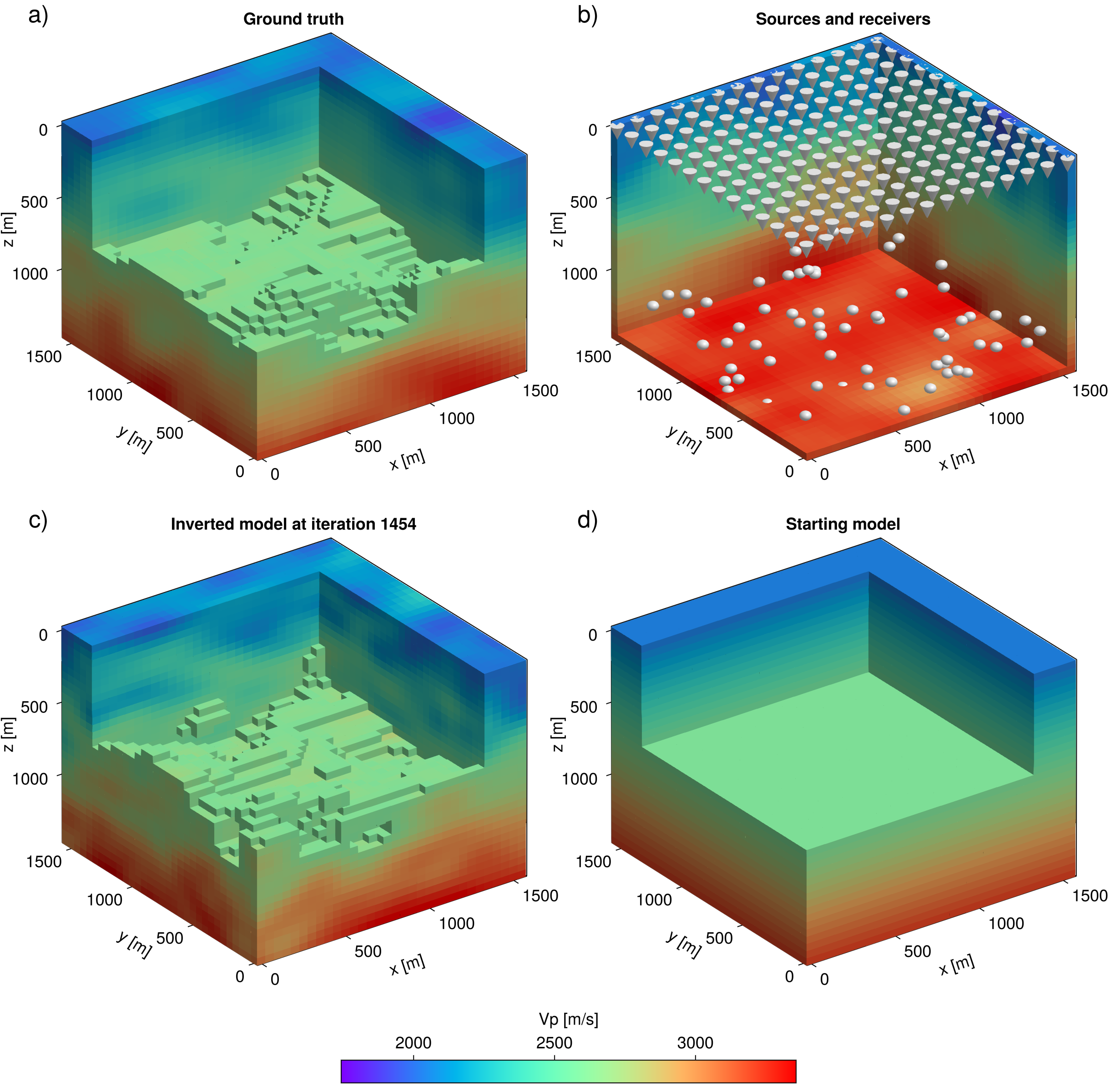}
  \caption{Probabilistic inversion for a 3D velocity model. a) Ground-truth model from which the observed data (with the addition of noise) have been generated. b) Source (spheres) and receiver (cone) locations. c) A random model from the posterior collection after 1454 iterations. d) Starting model, a laterally-homogeneous velocity model and random location of sources. .}
  \label{fig:hmc_vel3D}
\end{figure}

The second example concerns a 3D probabilistic inversion targeting the velocity structure. The size of the model is $32 \times  32 \times 30$ along $x$, $y$ and $z$, respectively, for a total of 30720 model parameters (Fig.~\ref{fig:hmc_vel3D}a). The 64 sources are located in the bottom part of the model and their location is assumed to be known, while the 128 receivers are all placed near the surface (Fig.~\ref{fig:hmc_vel3D}b). For this example, 10000 iterations with the NUTS algorithm were run. The data misfit decreases relatively rapidly, showcasing a short burn-in time. The resulting posterior models are in good agreement with the ground-truth model, although they appear to deviate more than for the 2D case, due to the smaller ray density. 

\section{Conclusions}
We have developed a formalism to perform eikonal traveltime tomography using the discrete adjoint method. The discrete adjoint method naturally handles boundary conditions and assures that the computed gradient corresponds to the discretised forward model. This approach allows us to efficiently compute gradients both with respect to velocity structure and source location within the same framework. Moreover, within our framework, sources and receivers position can be arbitrary, avoiding the need to approximate it to the nearest grid point. Refinement of the grid around the source to improve accuracy is also part of our methodology and is taken into account in the gradient calculations.
Using a (second order) FMM to solve the forward problem allows us to exploit some of its peculiar properties to solve the adjoint problem, which leads to the solution of a triangular linear system and therefore efficient computation of gradients.
We have demonstrated applications of this formalism to a set of 2D and 3D  synthetic tests both for deterministic and probabilistic inversion approaches. In the deterministic case, we have employed an l-BFGS optimization algorithm. For the probabilistic inversion examples, instead, we have employed a HMC algorithm, which requires the computation of gradients to efficiently sample the posterior probability density function. In both cases the algorithms produce good results. 
As expected, joint inversion for both velocity structure and source location is more challenging than targeting velocity alone.


\section{Data Availability}
The Julia code for forward and inverse calculations as performed in this paper is available as part of the G$^{-1}$Lab suite   \citep[former HMCLab,][]{zuninoHMCLabFrameworkSolving2023a} in a public repository at \url{https://github.com/GinvLab}.


\section{Acknowledgments}
All plots have been generated using the Julia package Makie \citep{danischMakieJlFlexible2021}.

\bibliographystyle{plainnat}
\bibliography{Traveltime_tomo}

\begin{thebibliography}{78}
\providecommand{\natexlab}[1]{#1}
\providecommand{\url}[1]{\texttt{#1}}
\expandafter\ifx\csname urlstyle\endcsname\relax
  \providecommand{\doi}[1]{doi: #1}\else
  \providecommand{\doi}{doi: \begingroup \urlstyle{rm}\Url}\fi

\bibitem[Aki and Lee(1976)]{akiDeterminationThreedimensionalVelocity1976}
Keiiti Aki and W.~H.~K. Lee.
\newblock Determination of three-dimensional velocity anomalies under a seismic
  array using first {{P}} arrival times from local earthquakes: 1. {{A}}
  homogeneous initial model.
\newblock \emph{Journal of Geophysical Research (1896-1977)}, 81\penalty0
  (23):\penalty0 4381--4399, 1976.
\newblock ISSN 2156-2202.
\newblock \doi{10.1029/JB081i023p04381}.

\bibitem[Aleardi and Salusti(2020)]{aleardiHamiltonianMonteCarlo2020a}
Mattia Aleardi and Alessandro Salusti.
\newblock Hamiltonian {{Monte Carlo}} algorithms for target- and
  interval-oriented amplitude versus angle inversions.
\newblock \emph{GEOPHYSICS}, 85\penalty0 (3):\penalty0 R177--R194, May 2020.
\newblock ISSN 0016-8033, 1942-2156.
\newblock \doi{10.1190/geo2019-0517.1}.

\bibitem[Benaichouche et~al.(2015)Benaichouche, Noble, and
  Gesret]{benaichoucheFirstArrivalTraveltime2015}
A.~Benaichouche, M.~Noble, and A.~Gesret.
\newblock First {{Arrival Traveltime Tomography Using}} the {{Fast Marching
  Method}} and the {{Adjoint State Technique}}.
\newblock In \emph{77th {{EAGE Conference}} and {{Exhibition}} 2015}, Madrid,
  Spain, June 2015.
\newblock \doi{10.3997/2214-4609.201412568}.

\bibitem[Betancourt(2017)]{betancourtConceptualIntroductionHamiltonian2017}
Michael Betancourt.
\newblock A {{Conceptual Introduction}} to {{Hamiltonian Monte Carlo}}.
\newblock \emph{arXiv:1701.02434 [stat]}, 2017.

\bibitem[Bezanson et~al.(2017)Bezanson, Edelman, Karpinski, and
  Shah]{bezansonJuliaFreshApproach2017}
Jeff Bezanson, Alan Edelman, Stefan Karpinski, and Viral~B. Shah.
\newblock Julia: {{A Fresh Approach}} to {{Numerical Computing}}.
\newblock \emph{SIAM Review}, 59\penalty0 (1):\penalty0 65--98, 2017.
\newblock ISSN 0036-1445.
\newblock \doi{10.1137/141000671}.

\bibitem[bin Waheed et~al.(2016)bin Waheed, Flagg, and
  Yarman]{waheedFirstarrivalTraveltimeTomography2016}
Umair bin Waheed, Garret Flagg, and Can~Evren Yarman.
\newblock First-arrival traveltime tomography for anisotropic media using the
  adjoint-state method.
\newblock \emph{GEOPHYSICS}, 81\penalty0 (4):\penalty0 R147--R155, July 2016.
\newblock ISSN 0016-8033, 1942-2156.
\newblock \doi{10.1190/geo2015-0463.1}.

\bibitem[Bozda{\u g} et~al.(2016)Bozda{\u g}, Peter, Lefebvre, Komatitsch,
  Tromp, Hill, Podhorszki, and Pugmire]{bozdagGlobalAdjointTomography2016}
Ebru Bozda{\u g}, Daniel Peter, Matthieu Lefebvre, Dimitri Komatitsch, Jeroen
  Tromp, Judith Hill, Norbert Podhorszki, and David Pugmire.
\newblock Global adjoint tomography: First-generation model.
\newblock \emph{Geophysical Journal International}, 207\penalty0 (3):\penalty0
  1739--1766, December 2016.
\newblock ISSN 0956-540X.
\newblock \doi{10.1093/gji/ggw356}.

\bibitem[Brossier et~al.(2009)Brossier, Operto, and Virieux]{Brossier_2009a}
R.~Brossier, S.~Operto, and J.~Virieux.
\newblock Seismic imaging of complex onshore structures by {{2D}} elastic
  frequency-domain full-waveform inversion.
\newblock \emph{Geophysics}, 74:\penalty0 WCC105--WCC118, 2009.

\bibitem[Bunks et~al.(1995)Bunks, Saleck, Zaleski, and
  Chavent]{bunksMultiscaleSeismicWaveform1995}
Carey Bunks, Fatimetou~M. Saleck, S.~Zaleski, and G.~Chavent.
\newblock Multiscale seismic waveform inversion.
\newblock \emph{Geophysics}, 60\penalty0 (5):\penalty0 1457--1473, October
  1995.
\newblock ISSN 0016-8033.
\newblock \doi{10.1190/1.1443880}.

\bibitem[Cerveny(2001)]{cervenySeismicRayTheory2001}
V.~Cerveny.
\newblock \emph{Seismic {{Ray Theory}}}.
\newblock Cambridge University Press, July 2001.
\newblock ISBN 978-0-521-36671-7.

\bibitem[Charara et~al.(2000)Charara, Barnes, and
  Tarantola]{chararaFullWaveformInversion2000}
Marwan Charara, Christophe Barnes, and Albert Tarantola.
\newblock Full waveform inversion of seismic data for a viscoelastic medium.
\newblock In \emph{Methods and Applications of Inversion}, pages 68--81.
  Springer, 2000.

\bibitem[Danisch and Krumbiegel(2021)]{danischMakieJlFlexible2021}
Simon Danisch and Julius Krumbiegel.
\newblock Makie.jl: {{Flexible}} high-performance data visualization for
  {{Julia}}.
\newblock \emph{Journal of Open Source Software}, 6\penalty0 (65):\penalty0
  3349, September 2021.
\newblock ISSN 2475-9066.
\newblock \doi{10.21105/joss.03349}.

\bibitem[Dhabaria and Singh(2024)]{dhabariaHamiltonianMonteCarlo2024}
Nirmit Dhabaria and Satish~C Singh.
\newblock Hamiltonian {{Monte Carlo}} based elastic full-waveform inversion of
  wide-angle seismic data.
\newblock \emph{Geophysical Journal International}, 237\penalty0 (3):\penalty0
  1384--1399, June 2024.
\newblock ISSN 1365-246X.
\newblock \doi{10.1093/gji/ggae112}.

\bibitem[Duane et~al.(1987)Duane, Kennedy, Pendleton, and
  Roweth]{duaneHybridMonteCarlo1987}
Simon Duane, A.~D. Kennedy, Brian~J. Pendleton, and Duncan Roweth.
\newblock Hybrid {{Monte Carlo}}.
\newblock \emph{Physics Letters B}, 195\penalty0 (2):\penalty0 216--222, 1987.
\newblock ISSN 0370-2693.
\newblock \doi{10.1016/0370-2693(87)91197-X}.

\bibitem[Fichtner(2011)]{fichtnerFullSeismicWaveform2011}
Andreas Fichtner.
\newblock \emph{Full {{Seismic Waveform Modelling}} and {{Inversion}}}.
\newblock Advances in {{Geophysical}} and {{Environmental Mechanics}} and
  {{Mathematics}}. Springer, Berlin, Heidelberg, 2011.
\newblock ISBN 978-3-642-15806-3 978-3-642-15807-0.
\newblock \doi{10.1007/978-3-642-15807-0}.

\bibitem[Fichtner et~al.(2019)Fichtner, Zunino, and
  Gebraad]{fichtnerHamiltonianMonteCarlo2019}
Andreas Fichtner, Andrea Zunino, and Lars Gebraad.
\newblock Hamiltonian {{Monte Carlo}} solution of tomographic inverse problems.
\newblock \emph{Geophysical Journal International}, 216\penalty0 (2):\penalty0
  1344--1363, 2019.
\newblock ISSN 0956-540X.
\newblock \doi{10.1093/gji/ggy496}.

\bibitem[Fomel et~al.(2009)Fomel, Luo, and Zhao]{fomelFastSweepingMethod2009}
Sergey Fomel, Songting Luo, and Hongkai Zhao.
\newblock Fast sweeping method for the factored eikonal equation.
\newblock \emph{Journal of Computational Physics}, 228\penalty0 (17):\penalty0
  6440--6455, September 2009.
\newblock ISSN 00219991.
\newblock \doi{10.1016/j.jcp.2009.05.029}.

\bibitem[Gebraad et~al.(2020)Gebraad, Boehm, and
  Fichtner]{gebraadBayesianElasticFullWaveform2020}
Lars Gebraad, Christian Boehm, and Andreas Fichtner.
\newblock Bayesian {{Elastic Full-Waveform Inversion Using Hamiltonian Monte
  Carlo}}.
\newblock \emph{Journal of Geophysical Research: Solid Earth}, 125\penalty0
  (3):\penalty0 e2019JB018428, 2020.
\newblock ISSN 2169-9356.
\newblock \doi{10.1029/2019JB018428}.

\bibitem[{Gerhard Pratt} et~al.(1998){Gerhard Pratt}, Shin, and
  {Hicks}]{gerhardprattGaussNewtonFullNewton1998}
{Gerhard Pratt}, Changsoo Shin, and {Hicks}.
\newblock Gauss-{{Newton}} and full {{Newton}} methods in frequency-space
  seismic waveform inversion.
\newblock \emph{Geophysical Journal International}, 133\penalty0 (2):\penalty0
  341--362, May 1998.
\newblock ISSN 0956540X, 1365246X.
\newblock \doi{10.1046/j.1365-246X.1998.00498.x}.

\bibitem[Girolami and Calderhead(2011)]{girolamiRiemannManifoldLangevin2011}
Mark Girolami and Ben Calderhead.
\newblock Riemann manifold {{Langevin}} and {{Hamiltonian Monte Carlo}}
  methods: {{Riemann Manifold Langevin}} and {{Hamiltonian Monte Carlo
  Methods}}.
\newblock \emph{Journal of the Royal Statistical Society: Series B (Statistical
  Methodology)}, 73\penalty0 (2):\penalty0 123--214, March 2011.
\newblock ISSN 13697412.
\newblock \doi{10.1111/j.1467-9868.2010.00765.x}.

\bibitem[Hastings(1970)]{hastingsMonteCarloSampling1970}
W.~K. Hastings.
\newblock Monte {{Carlo}} sampling methods using {{Markov}} chains and their
  applications.
\newblock \emph{Biometrika}, 57\penalty0 (1):\penalty0 97--109, April 1970.
\newblock ISSN 0006-3444.
\newblock \doi{10.1093/biomet/57.1.97}.

\bibitem[Hinze et~al.(2008)Hinze, Pinnau, Ulbrich, and
  Ulbrich]{hinzeOptimizationPDEConstraints2008}
Michael Hinze, Rene Pinnau, Michael Ulbrich, and Stefan Ulbrich.
\newblock \emph{Optimization with {{PDE Constraints}}}.
\newblock Springer, New York, 2009th edition edition, November 2008.
\newblock ISBN 978-1-4020-8838-4.

\bibitem[Hoffman and Gelman(2014)]{hoffmanNoUturnSamplerAdaptively2014}
Matthew~D. Hoffman and Andrew Gelman.
\newblock The {{No-U-turn}} sampler: Adaptively setting path lengths in
  {{Hamiltonian Monte Carlo}}.
\newblock \emph{The Journal of Machine Learning Research}, 15\penalty0
  (1):\penalty0 1593--1623, 2014.
\newblock ISSN 1532-4435.

\bibitem[Huang and Bellefleur(2012)]{huangJointTransmissionReflection2012}
Jun-Wei Huang and Gilles Bellefleur.
\newblock Joint transmission and reflection traveltime tomography using the
  fast sweeping method and the adjoint-state technique: {{Joint}} traveltime
  tomography.
\newblock \emph{Geophysical Journal International}, 188\penalty0 (2):\penalty0
  570--582, February 2012.
\newblock ISSN 0956540X.
\newblock \doi{10.1111/j.1365-246X.2011.05273.x}.

\bibitem[Husen and Kissling(2001)]{husenLocalEarthquakeTomography2001}
Stephan Husen and Edi Kissling.
\newblock Local earthquake tomography between rays and waves: Fat ray
  tomography.
\newblock \emph{Physics of the Earth and Planetary Interiors}, 123\penalty0
  (2):\penalty0 127--147, April 2001.
\newblock ISSN 0031-9201.
\newblock \doi{10.1016/S0031-9201(00)00206-5}.

\bibitem[Kissling et~al.(1994)Kissling, Ellsworth, {Eberhart-Phillips}, and
  Kradolfer]{kisslingInitialReferenceModels1994}
E.~Kissling, W.~L. Ellsworth, D.~{Eberhart-Phillips}, and U.~Kradolfer.
\newblock Initial reference models in local earthquake tomography.
\newblock \emph{Journal of Geophysical Research: Solid Earth}, 99\penalty0
  (B10):\penalty0 19635--19646, 1994.
\newblock ISSN 2156-2202.
\newblock \doi{10.1029/93JB03138}.

\bibitem[Le~Bouteiller et~al.(2019)Le~Bouteiller, Benjemaa, M{\'e}tivier, and
  Virieux]{lebouteillerDiscontinuousGalerkinFastsweeping2019}
Philippe Le~Bouteiller, Mondher Benjemaa, Ludovic M{\'e}tivier, and Jean
  Virieux.
\newblock A discontinuous {{Galerkin}} fast-sweeping eikonal solver for fast
  and accurate traveltime computation in {{3D}} tilted anisotropic media.
\newblock \emph{GEOPHYSICS}, 84\penalty0 (2):\penalty0 C107--C118, March 2019.
\newblock ISSN 0016-8033, 1942-2156.
\newblock \doi{10.1190/geo2018-0555.1}.

\bibitem[Leli{\`e}vre et~al.(2011)Leli{\`e}vre, Farquharson, and
  Hurich]{lelievreInversionFirstarrivalSeismic2011}
Peter~G. Leli{\`e}vre, Colin~G. Farquharson, and Charles~A. Hurich.
\newblock Inversion of first-arrival seismic traveltimes without rays,
  implemented on unstructured grids: {{Inversion}} of first-arrivals without
  rays.
\newblock \emph{Geophysical Journal International}, 185\penalty0 (2):\penalty0
  749--763, May 2011.
\newblock ISSN 0956540X.
\newblock \doi{10.1111/j.1365-246X.2011.04964.x}.

\bibitem[Leung and Qian(2006)]{leungAdjointStateMethod2006a}
Shingyu Leung and Jianliang Qian.
\newblock An adjoint state method for three-dimensional transmission traveltime
  tomography using first-arrivals.
\newblock \emph{Communications in Mathematical Sciences}, 4\penalty0
  (1):\penalty0 249--266, March 2006.
\newblock ISSN 1945-0796.
\newblock \doi{10.4310/CMS.2006.v4.n1.a10}.

\bibitem[Li et~al.(2013)Li, Vladimirsky, and
  Fomel]{liFirstbreakTraveltimeTomography2013a}
Siwei Li, Alexander Vladimirsky, and Sergey Fomel.
\newblock First-break traveltime tomography with the double-square-root eikonal
  equation.
\newblock \emph{GEOPHYSICS}, 78\penalty0 (6):\penalty0 U89--U101, November
  2013.
\newblock ISSN 0016-8033.
\newblock \doi{10.1190/geo2013-0058.1}.

\bibitem[Li et~al.(2014)Li, Leung, and Qian]{liLevelsetAdjointstateMethod2014}
Wenbin Li, Shingyu Leung, and Jianliang Qian.
\newblock A level-set adjoint-state method for crosswell
  transmission-reflection traveltime tomography.
\newblock \emph{Geophysical Journal International}, 199\penalty0 (1):\penalty0
  348--367, October 2014.
\newblock ISSN 1365-246X, 0956-540X.
\newblock \doi{10.1093/gji/ggu262}.

\bibitem[Lions(1971)]{lionsOptimalControlSystems1971}
Jacques-Louis Lions.
\newblock \emph{Optimal {{Control}} of {{Systems Governed}} by {{Partial
  Differential Equations}}}.
\newblock Springer-Verlag, 1971.
\newblock ISBN 978-0-387-05115-4.

\bibitem[Liu and Nocedal(1989)]{liuLimitedMemoryBFGS1989}
Dong~C. Liu and Jorge Nocedal.
\newblock On the limited memory {{BFGS}} method for large scale optimization.
\newblock \emph{Mathematical Programming}, 45\penalty0 (1):\penalty0 503--528,
  August 1989.
\newblock ISSN 1436-4646.
\newblock \doi{10.1007/BF01589116}.

\bibitem[Luo et~al.(2012)Luo, Qian, and Zhao]{luoHigherorderSchemes3D2012a}
Songting Luo, Jianliang Qian, and Hongkai Zhao.
\newblock Higher-order schemes for {{3D}} first-arrival traveltimes and
  amplitudes.
\newblock \emph{GEOPHYSICS}, 77\penalty0 (2):\penalty0 T47--T56, March 2012.
\newblock ISSN 0016-8033.
\newblock \doi{10.1190/geo2010-0363.1}.

\bibitem[Metropolis et~al.(1953)Metropolis, Rosenbluth, Rosenbluth, Teller, and
  Teller]{metropolisEquationStateCalculations1953b}
Nicholas Metropolis, Arianna~W. Rosenbluth, Marshall~N. Rosenbluth, Augusta~H.
  Teller, and Edward Teller.
\newblock Equation of {{State Calculations}} by {{Fast Computing Machines}}.
\newblock \emph{The Journal of Chemical Physics}, 21\penalty0 (6):\penalty0
  1087--1092, June 1953.
\newblock ISSN 0021-9606.
\newblock \doi{10.1063/1.1699114}.

\bibitem[Mosegaard and Tarantola(2002)]{mosegaard16ProbabilisticApproach2002}
Klaus Mosegaard and Albert Tarantola.
\newblock 16 {{Probabilistic}} approach to inverse problems.
\newblock In \emph{International {{Geophysics}}}, volume~81, pages 237--265.
  Elsevier, 2002.
\newblock ISBN 978-0-12-440652-0.
\newblock \doi{10.1016/S0074-6142(02)80219-4}.

\bibitem[Neal(2011)]{nealMCMCUsingHamiltonian2011}
Radford~M. Neal.
\newblock {{MCMC Using Hamiltonian Dynamics}}.
\newblock In \emph{Handbook of {{Markov Chain Monte Carlo}}}. {Chapman and
  Hall/CRC}, 2011.
\newblock ISBN 978-0-429-13850-8.

\bibitem[Noble et~al.(2014)Noble, Gesret, and
  Belayouni]{nobleAccurate3DFinite2014}
M.~Noble, A.~Gesret, and N.~Belayouni.
\newblock Accurate 3-{{D}} finite difference computation of traveltimes in
  strongly heterogeneous media.
\newblock \emph{Geophysical Journal International}, 199\penalty0 (3):\penalty0
  1572--1585, December 2014.
\newblock ISSN 0956-540X.
\newblock \doi{10.1093/gji/ggu358}.

\bibitem[Nocedal(1980)]{nocedalUpdatingQuasiNewtonMatrices1980}
Jorge Nocedal.
\newblock Updating quasi-{{Newton}} matrices with limited storage.
\newblock \emph{Mathematics of Computation}, 35\penalty0 (151):\penalty0
  773--782, 1980.
\newblock ISSN 0025-5718, 1088-6842.
\newblock \doi{10.1090/S0025-5718-1980-0572855-7}.

\bibitem[Nocedal and Wright(2006)]{nocedalNumericalOptimization2006}
Jorge Nocedal and Stephen~J. Wright.
\newblock \emph{Numerical Optimization}.
\newblock Springer Series in Operations Research. Springer, New York, 2nd ed
  edition, 2006.
\newblock ISBN 978-0-387-30303-1.

\bibitem[Nolet(2008)]{noletBreviarySeismicTomography2008}
G.~Nolet.
\newblock \emph{A {{Breviary}} of {{Seismic Tomography}}}.
\newblock Cambridge University Press, September 2008.

\bibitem[Pica(1997)]{picaFastAccurateFinite1997}
A.~Pica.
\newblock Fast and accurate finite-difference solutions of the {{3D}} eikonal
  equation parametrized in celerity.
\newblock In \emph{{{SEG Technical Program Expanded Abstracts}} 1997}, pages
  1774--1777. Society of Exploration Geophysicists, January 1997.
\newblock \doi{10.1190/1.1885777}.

\bibitem[Plessix(2006)]{plessixReviewAdjointstateMethod2006b}
R.-E. Plessix.
\newblock A review of the adjoint-state method for computing the gradient of a
  functional with geophysical applications.
\newblock \emph{Geophysical Journal International}, 167\penalty0 (2):\penalty0
  495--503, November 2006.
\newblock ISSN 0956540X, 1365246X.
\newblock \doi{10.1111/j.1365-246X.2006.02978.x}.

\bibitem[Podvin and Lecomte(1991)]{podvinFiniteDifferenceComputation1991}
Pascal Podvin and Isabelle Lecomte.
\newblock Finite difference computation of traveltimes in very contrasted
  velocity models: A massively parallel approach and its associated tools.
\newblock \emph{Geophysical Journal International}, 105\penalty0 (1):\penalty0
  271--284, April 1991.
\newblock ISSN 0956540X, 1365246X.
\newblock \doi{10.1111/j.1365-246X.1991.tb03461.x}.

\bibitem[Qian and Symes(2002)]{qianAdaptiveFiniteDifference2002}
Jianliang Qian and William~W. Symes.
\newblock An adaptive finite-difference method for traveltimes and amplitudes.
\newblock \emph{GEOPHYSICS}, 67\penalty0 (1):\penalty0 167--176, January 2002.
\newblock ISSN 0016-8033.
\newblock \doi{10.1190/1.1451472}.

\bibitem[Rawlinson and
  Sambridge(2003)]{rawlinsonSeismicTraveltimeTomography2003}
N.~Rawlinson and M.~Sambridge.
\newblock Seismic traveltime tomography of the crust and the lithosphere.
\newblock In \emph{Advances in {{Geophysics}}}, volume~46, pages 81--198.
  Elsevier, January 2003.
\newblock \doi{10.1016/S0065-2687(03)46002-0}.

\bibitem[Rawlinson and
  Sambridge(2004{\natexlab{a}})]{rawlinsonMultipleReflectionTransmission2004}
N.~Rawlinson and M.~Sambridge.
\newblock Multiple reflection and transmission phases in complex layered media
  using a multistage fast marching method.
\newblock \emph{GEOPHYSICS}, 69\penalty0 (5):\penalty0 1338--1350, September
  2004{\natexlab{a}}.
\newblock ISSN 0016-8033, 1942-2156.
\newblock \doi{10.1190/1.1801950}.

\bibitem[Rawlinson and
  Sambridge(2004{\natexlab{b}})]{rawlinsonWaveFrontEvolution2004}
N.~Rawlinson and M.~Sambridge.
\newblock Wave front evolution in strongly heterogeneous layered media using
  the fast marching method.
\newblock \emph{Geophysical Journal International}, 156\penalty0 (3):\penalty0
  631--647, March 2004{\natexlab{b}}.
\newblock ISSN 0956540X, 1365246X.
\newblock \doi{10.1111/j.1365-246X.2004.02153.x}.

\bibitem[Rawlinson et~al.(2010)Rawlinson, Pozgay, and
  Fishwick]{rawlinsonSeismicTomographyWindow2010}
N.~Rawlinson, S.~Pozgay, and S.~Fishwick.
\newblock Seismic tomography: {{A}} window into deep {{Earth}}.
\newblock \emph{Physics of the Earth and Planetary Interiors}, 178\penalty0
  (3):\penalty0 101--135, February 2010.
\newblock ISSN 0031-9201.
\newblock \doi{10.1016/j.pepi.2009.10.002}.

\bibitem[Sei and Symes(1994)]{seiGradientCalculationTraveltime1994}
Alain Sei and William~W. Symes.
\newblock Gradient calculation of the traveltime cost function without ray
  tracing.
\newblock In \emph{{{SEG Technical Program Expanded Abstracts}} 1994}, {{SEG
  Technical Program Expanded Abstracts}}, pages 1351--1354. Society of
  Exploration Geophysicists, January 1994.
\newblock \doi{10.1190/1.1822780}.

\bibitem[Sen and Biswas(2017)]{senTransdimensionalSeismicInversion2017}
Mrinal~K. Sen and Reetam Biswas.
\newblock Transdimensional seismic inversion using the reversible jump
  {{Hamiltonian Monte Carlo}} algorithm.
\newblock \emph{Geophysics}, 82\penalty0 (3):\penalty0 R119--R134, 2017.
\newblock ISSN 0016-8033.
\newblock \doi{10.1190/geo2016-0010.1}.

\bibitem[Sethian(1996)]{sethianFastMarchingLevel1996a}
J~A Sethian.
\newblock A fast marching level set method for monotonically advancing fronts.
\newblock \emph{Proceedings of the National Academy of Sciences}, 93\penalty0
  (4):\penalty0 1591--1595, February 1996.
\newblock \doi{10.1073/pnas.93.4.1591}.

\bibitem[Sethian(1999{\natexlab{a}})]{sethianAdvancesFastMarching1999}
J.~A. Sethian.
\newblock Advances in {{Fast Marching}} and {{Level Set Methods}} for
  {{Propagating Interfaces}}.
\newblock In Rolf Jeltsch and Michael Fey, editors, \emph{Hyperbolic
  {{Problems}}: {{Theory}}, {{Numerics}}, {{Applications}}}, pages 855--864.
  Birkh{\"a}user Basel, Basel, 1999{\natexlab{a}}.
\newblock ISBN 978-3-0348-9744-0 978-3-0348-8724-3.
\newblock \doi{10.1007/978-3-0348-8724-3_36}.

\bibitem[Sethian(1999{\natexlab{b}})]{sethianFastMarchingMethods1999}
J.~A. Sethian.
\newblock Fast {{Marching Methods}}.
\newblock \emph{SIAM Review}, 41\penalty0 (2):\penalty0 199--235, January
  1999{\natexlab{b}}.
\newblock ISSN 0036-1445.
\newblock \doi{10.1137/S0036144598347059}.

\bibitem[Sethian and Popovici(1999)]{sethian3DTraveltimeComputation1999}
James~A. Sethian and A.~Mihai Popovici.
\newblock 3-{{D}} traveltime computation using the fast marching method.
\newblock \emph{GEOPHYSICS}, 64\penalty0 (2):\penalty0 516--523, March 1999.
\newblock ISSN 0016-8033, 1942-2156.
\newblock \doi{10.1190/1.1444558}.

\bibitem[Simmons et~al.(2012)Simmons, Myers, Johannesson, and
  Matzel]{simmonsLLNLG3Dv3GlobalWave2012}
N.~A. Simmons, S.~C. Myers, G.~Johannesson, and E.~Matzel.
\newblock {{LLNL-G3Dv3}}: {{Global P}} wave tomography model for improved
  regional and teleseismic travel time prediction.
\newblock \emph{Journal of Geophysical Research: Solid Earth}, 117\penalty0
  (B10), 2012.
\newblock ISSN 2156-2202.
\newblock \doi{10.1029/2012JB009525}.

\bibitem[Taillandier et~al.(2009)Taillandier, Noble, Chauris, and
  Calandra]{taillandierFirstarrivalTraveltimeTomography2009}
C{\'e}dric Taillandier, Mark Noble, Herv{\'e} Chauris, and Henri Calandra.
\newblock First-arrival traveltime tomography based on the adjoint-state
  method.
\newblock \emph{Geophysics}, 74\penalty0 (6):\penalty0 WCB1--WCB10, November
  2009.
\newblock ISSN 0016-8033, 1942-2156.
\newblock \doi{10.1190/1.3250266}.

\bibitem[Tape et~al.(2009)Tape, Liu, Maggi, and
  Tromp]{tapeAdjointTomographySouthern2009}
Carl Tape, Qinya Liu, Alessia Maggi, and Jeroen Tromp.
\newblock Adjoint {{Tomography}} of the {{Southern California Crust}}.
\newblock \emph{Science}, 325\penalty0 (5943):\penalty0 988--992, August 2009.
\newblock \doi{10.1126/science.1175298}.

\bibitem[Tarantola(2005)]{tarantolaInverseProblemTheory2005}
Albert Tarantola.
\newblock \emph{Inverse {{Problem Theory}} and {{Methods}} for {{Model
  Parameter Estimation}}}.
\newblock Other {{Titles}} in {{Applied Mathematics}}. {Society for Industrial
  and Applied Mathematics}, January 2005.
\newblock ISBN 978-0-89871-572-9.
\newblock \doi{10.1137/1.9780898717921}.

\bibitem[Tavakoli~B. et~al.(2017)Tavakoli~B., Operto, Ribodetti, and
  Virieux]{tavakolib.SlopeTomographyBased2017}
F.~Tavakoli~B., S.~Operto, A.~Ribodetti, and J.~Virieux.
\newblock Slope tomography based on eikonal solvers and the adjoint-state
  method.
\newblock \emph{Geophysical Journal International}, 209\penalty0 (3):\penalty0
  1629--1647, June 2017.
\newblock ISSN 0956-540X, 1365-246X.
\newblock \doi{10.1093/gji/ggx111}.

\bibitem[Thurber(1983)]{thurberEarthquakeLocationsThreedimensional1983}
Clifford~H. Thurber.
\newblock Earthquake locations and three-dimensional crustal structure in the
  {{Coyote Lake Area}}, central {{California}}.
\newblock \emph{Journal of Geophysical Research: Solid Earth}, 88\penalty0
  (B10):\penalty0 8226--8236, 1983.
\newblock ISSN 2156-2202.
\newblock \doi{10.1029/JB088iB10p08226}.

\bibitem[Thurber(1992)]{thurberHypocentervelocityStructureCoupling1992}
Clifford~H. Thurber.
\newblock Hypocenter-velocity structure coupling in local earthquake
  tomography.
\newblock \emph{Physics of the Earth and Planetary Interiors}, 75\penalty0
  (1):\penalty0 55--62, December 1992.
\newblock ISSN 0031-9201.
\newblock \doi{10.1016/0031-9201(92)90117-E}.

\bibitem[Tong(2021{\natexlab{a}})]{tongAdjointStateTraveltimeTomography2021}
Ping Tong.
\newblock Adjoint-{{State Traveltime Tomography}}: {{Eikonal Equation-Based
  Methods}} and {{Application}} to the {{Anza Area}} in {{Southern
  California}}.
\newblock \emph{Journal of Geophysical Research: Solid Earth}, 126\penalty0
  (5):\penalty0 e2021JB021818, 2021{\natexlab{a}}.
\newblock ISSN 2169-9356.
\newblock \doi{10.1029/2021JB021818}.

\bibitem[Tong(2021{\natexlab{b}})]{tongAdjointStateTraveltimeTomography2021a}
Ping Tong.
\newblock Adjoint-{{State Traveltime Tomography}} for {{Azimuthally Anisotropic
  Media}} and {{Insight Into}} the {{Crustal Structure}} of {{Central
  California Near Parkfield}}.
\newblock \emph{Journal of Geophysical Research: Solid Earth}, 126\penalty0
  (10):\penalty0 e2021JB022365, 2021{\natexlab{b}}.
\newblock ISSN 2169-9356.
\newblock \doi{10.1029/2021JB022365}.

\bibitem[Tong et~al.(2024)Tong, Li, Chen, and
  Nagaso]{tongAdjointstateDifferentialArrival2024}
Ping Tong, Tianjue Li, Jing Chen, and Masaru Nagaso.
\newblock Adjoint-state differential arrival time tomography.
\newblock \emph{Geophysical Journal International}, 236\penalty0 (1):\penalty0
  139--160, January 2024.
\newblock ISSN 0956-540X.
\newblock \doi{10.1093/gji/ggad416}.

\bibitem[Treister and Haber(2016)]{treisterFastMarchingAlgorithm2016b}
Eran Treister and Eldad Haber.
\newblock A fast marching algorithm for the factored eikonal equation.
\newblock \emph{Journal of Computational Physics}, 324:\penalty0 210--225,
  November 2016.
\newblock ISSN 0021-9991.
\newblock \doi{10.1016/j.jcp.2016.08.012}.

\bibitem[Tromp et~al.(2004)Tromp, Tape, and
  Liu]{trompSeismicTomographyAdjoint2004}
Jeroen Tromp, Carl Tape, and Qinya Liu.
\newblock Seismic tomography, adjoint methods, time reversal and
  banana-doughnut kernels: {{Seismic}} tomography, adjoint methods, time
  reversal and banana-doughnut kernels.
\newblock \emph{Geophysical Journal International}, 160\penalty0 (1):\penalty0
  195--216, December 2004.
\newblock ISSN 0956540X, 1365246X.
\newblock \doi{10.1111/j.1365-246X.2004.02453.x}.

\bibitem[van Trier and Symes(1991)]{trierUpwindFinitedifferenceCalculation1991}
Jos van Trier and William~W. Symes.
\newblock Upwind finite-difference calculation of traveltimes.
\newblock \emph{Geophysics}, 56\penalty0 (6):\penalty0 812--821, June 1991.
\newblock ISSN 0016-8033.
\newblock \doi{10.1190/1.1443099}.

\bibitem[Vidale(1988)]{vidaleFinitedifferenceCalculationTravel1988}
J.~Vidale.
\newblock Finite-difference calculation of travel times.
\newblock \emph{Bulletin of the Seismological Society of America}, 1988.

\bibitem[Vidale(1990)]{vidaleFinitedifferenceCalculationTraveltimes1990}
John~E. Vidale.
\newblock Finite-difference calculation of traveltimes in three dimensions.
\newblock \emph{Geophysics}, 55\penalty0 (5):\penalty0 521--526, May 1990.
\newblock ISSN 0016-8033.
\newblock \doi{10.1190/1.1442863}.

\bibitem[Virieux and Operto(2009)]{virieuxOverviewFullwaveformInversion2009}
J.~Virieux and S.~Operto.
\newblock An overview of full-waveform inversion in exploration geophysics.
\newblock \emph{GEOPHYSICS}, 74\penalty0 (6):\penalty0 WCC1--WCC26, November
  2009.
\newblock ISSN 0016-8033.
\newblock \doi{10.1190/1.3238367}.

\bibitem[Virieux and Farra(1991)]{virieuxRayTracing3D1991}
Jean Virieux and Veronique Farra.
\newblock Ray tracing in 3-{{D}} complex isotropic media; an analysis of the
  problem.
\newblock \emph{Geophysics}, 56\penalty0 (12):\penalty0 2057--2069, December
  1991.
\newblock ISSN 0016-8033.
\newblock \doi{10.1190/1.1443018}.

\bibitem[Zelt and Barton(1998)]{zeltThreedimensionalSeismicRefraction1998}
Colin~A. Zelt and Penny~J. Barton.
\newblock Three-dimensional seismic refraction tomography: {{A}} comparison of
  two methods applied to data from the {{Faeroe Basin}}.
\newblock \emph{Journal of Geophysical Research: Solid Earth}, 103\penalty0
  (B4):\penalty0 7187--7210, 1998.
\newblock ISSN 2156-2202.
\newblock \doi{10.1029/97JB03536}.

\bibitem[Zhang et~al.(2005)Zhang, Rector, and
  Hoversten]{zhangEikonalSolverCelerity2005}
Linbin Zhang, James~W. Rector, and G.~Michael Hoversten.
\newblock Eikonal solver in the celerity domain.
\newblock \emph{Geophysical Journal International}, 162\penalty0 (1):\penalty0
  1--8, July 2005.
\newblock ISSN 0956540X, 1365246X.
\newblock \doi{10.1111/j.1365-246X.2005.02626.x}.

\bibitem[Zhao(2005)]{zhaoFastSweepingMethod2005}
Hongkai Zhao.
\newblock A fast sweeping method for {{Eikonal}} equations.
\newblock \emph{Mathematics of Computation}, 74\penalty0 (250):\penalty0
  603--627, April 2005.
\newblock ISSN 0025-5718, 1088-6842.
\newblock \doi{10.1090/S0025-5718-04-01678-3}.

\bibitem[Zunino and Mosegaard(2018)]{zuninoIntegratingGradientInformation2018}
A.~Zunino and K.~Mosegaard.
\newblock Integrating {{Gradient Information}} with {{Probabilistic Traveltime
  Tomography Using}} the {{Hamiltonian Monte Carlo Algorithm}}.
\newblock In \emph{80th {{EAGE Conference}} \& {{Exhibition}} 2018 {{Workshop
  Programme}}}, page~cp. European Association of Geoscientists \& Engineers,
  June 2018.
\newblock ISBN 978-94-6282-257-3.
\newblock \doi{10.3997/2214-4609.201801971}.

\bibitem[Zunino et~al.(2022)Zunino, Ghirotto, Armadillo, and
  Fichtner]{zuninoHamiltonianMonteCarlo2022a}
Andrea Zunino, Alessandro Ghirotto, Egidio Armadillo, and Andreas Fichtner.
\newblock Hamiltonian {{Monte Carlo Probabilistic Joint Inversion}} of {{2D}}
  (2.{{75D}}) {{Gravity}} and {{Magnetic Data}}.
\newblock \emph{Geophysical Research Letters}, 49\penalty0 (20):\penalty0
  e2022GL099789, 2022.
\newblock ISSN 1944-8007.
\newblock \doi{10.1029/2022GL099789}.

\bibitem[Zunino et~al.(2023)Zunino, Gebraad, Ghirotto, and
  Fichtner]{zuninoHMCLabFrameworkSolving2023a}
Andrea Zunino, Lars Gebraad, Alessandro Ghirotto, and Andreas Fichtner.
\newblock {{HMCLab}}: A framework for solving diverse geophysical inverse
  problems using the {{Hamiltonian Monte Carlo}} method.
\newblock \emph{Geophysical Journal International}, 235\penalty0 (3):\penalty0
  2979--2991, December 2023.
\newblock ISSN 0956-540X.
\newblock \doi{10.1093/gji/ggad403}.

\end{thebibliography}

\end{document}